\documentclass[aps,prl,twocolumn,showpacs,balance,amssymb,superscriptaddress]{revtex4-1} 

\usepackage{amsmath,amssymb,amsfonts,amsthm}   
\usepackage{graphicx} 											
\usepackage{hyperref}                                         
\usepackage{natbib}
\usepackage{listings}
\usepackage{bm}
\usepackage{dsfont}
\usepackage[ansinew]{inputenc}
\usepackage[ngerman,english]{babel}
\usepackage{color}
\usepackage{enumerate}

\newcommand{\hatbm}[1]{\hat{\bm{#1}}}

\newcommand{\mathsym}[1]{{}}
\newcommand{\unicode}[1]{{}}

\newcommand{\absa}{\vert \bm{k}_1-\bm{q}\vert}
\newcommand{\absb}{\vert \bm{k}_2+\bm{q}\vert}

\begin{document}

\title{Current amplification and relaxation in Dirac systems}
\author{Alexandra Junck}
\affiliation{\mbox{Dahlem Center for Complex Quantum Systems and Fachbereich Physik, Freie Universit\"at Berlin, 14195 Berlin, Germany}}
\author{Gil Refael}
\affiliation{\mbox{Department of Physics, California Institute of Technology, Pasadena, CA 91125 }}
\author{Felix von Oppen}
\affiliation{\mbox{Dahlem Center for Complex Quantum Systems and Fachbereich Physik, Freie Universit\"at Berlin, 14195 Berlin, Germany}}
\date{\today}
\begin{abstract}
We study how electron-electron (e-e) interactions affect the current carried by excited electrons and holes in Dirac systems such as graphene or topological insulators. We find that the current shows distinctly different behavior for e-e collisions involving the electron or the hole. Most surprisingly, for positive Fermi energy, collisions of the electrons can substantially increase the current. This remarkable amplification of the current can be of the order of $\sim 10\%$ per scattering event. It also causes a strong suppression of the overall current relaxation rate of a photoexcited electron-hole pair, with the amplification of the electron current making up for a fast decay of the hole current.
\end{abstract}
\maketitle

{\em Introduction.---}Dirac systems like graphene and topological insulators (TIs) are a central research topic in condensed matter physics. The easy fabrication process of graphene \cite{novoselov2005} and the discovery of a variety of materials that are 2D \cite{kane2005,konig2007,castro2009} or 3D \cite{fu2007,hsieh2008,hasan2010,qi2011} TIs are but two reasons why Dirac systems are studied intensively. Because of their linear dispersion and unique helical (pseudo)spin structure, graphene and TIs might be valuable materials for spintronic devices \cite{macdonald2012}. In addition, graphene and TIs are promising systems for applications in the rising field of optoelectronics, for instance as transparent conductors or photodetectors \cite{bonaccorso2010,kong2010}.

\begin{figure}[t]
 \includegraphics[width=8cm]{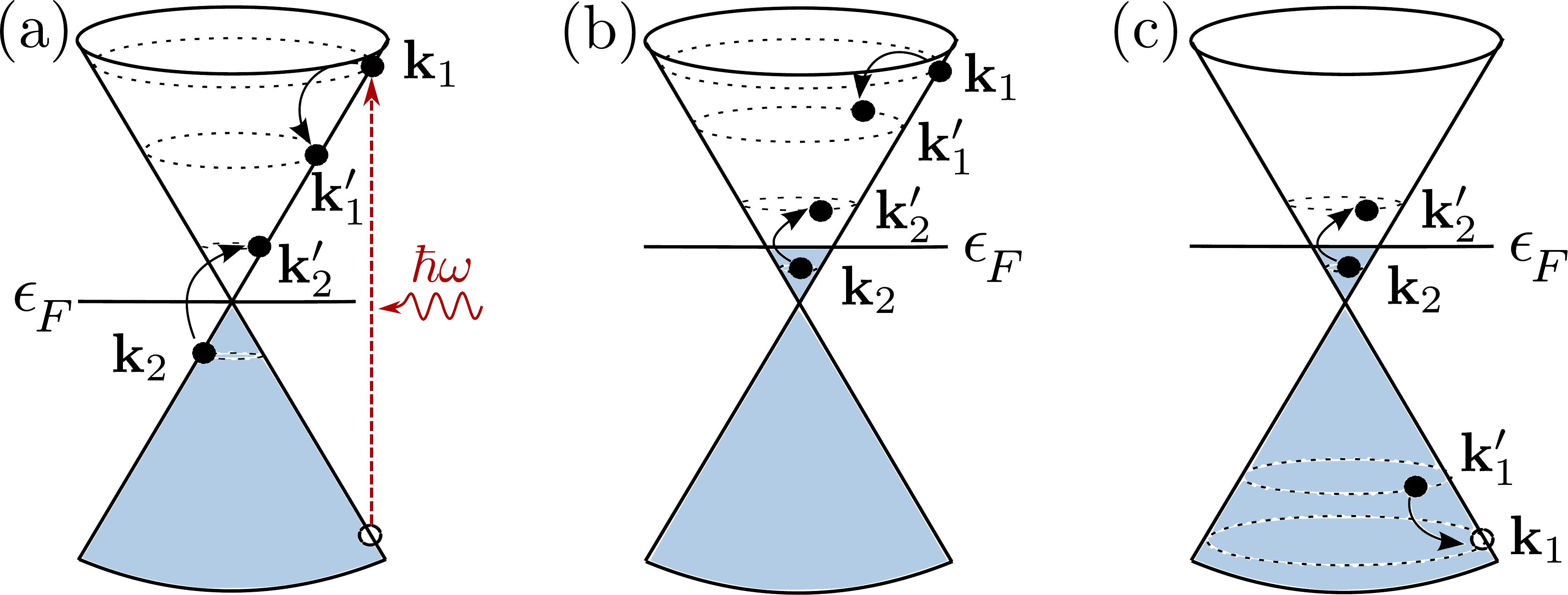}
 \caption{The photoexcitation of an electron-hole pair within the Dirac cone and the relaxation process of the hot electron by excitation of an electron-hole pair, for (a) $\epsilon_F=0$ and (b) $\epsilon_F >0$. (c) A possible relaxation process of the excited hole.}
 \label{fig:relax1}
\end{figure}

Photocurrents provide an interesting probe of the optoelectronic properties of Dirac materials, and have been measured in both systems \cite{park2009,mciver2011,kastl2012,sun2012,freitag2013}. Their magnitude is governed by a competition between carrier excitation which is asymmetric in momentum space and current relaxation \cite{junck2013}. While there has been extensive research on energy relaxation of excited electrons both theoretically \cite{cheianov2006,butscher2007,stauber2007,tse2009,winzer2010,kim2011,song2013} and experimentally \cite{hsieh2007,kumar2011,sobota2012,hajlaoui2012,gierz2013}, current relaxation of photoexcited carriers in Dirac systems remains much less unexplored \cite{sun2012}.
In general, current relaxation occurs through impurity, electron-phonon, or electron-electron (e-e) scattering. Here we study highly excited electrons, holes, or electron-hole pairs in the Dirac cone and assume that e-e interactions provide the dominant relaxation mechanism [see Fig.~\ref{fig:relax1}(a)]. Interestingly, e-e scattering does not contribute to current relaxation when the carrier dispersion is quadratic. Indeed, for quadratic dispersions velocity is proportional to momentum and thus momentum conservation implies current conservation. In contrast, in Dirac systems with their linear dispersion velocity is no longer proportional to momentum and the current can change and relax by e-e scattering.

\begin{figure}[t]
\begin{center}
\includegraphics[width=8.5cm]{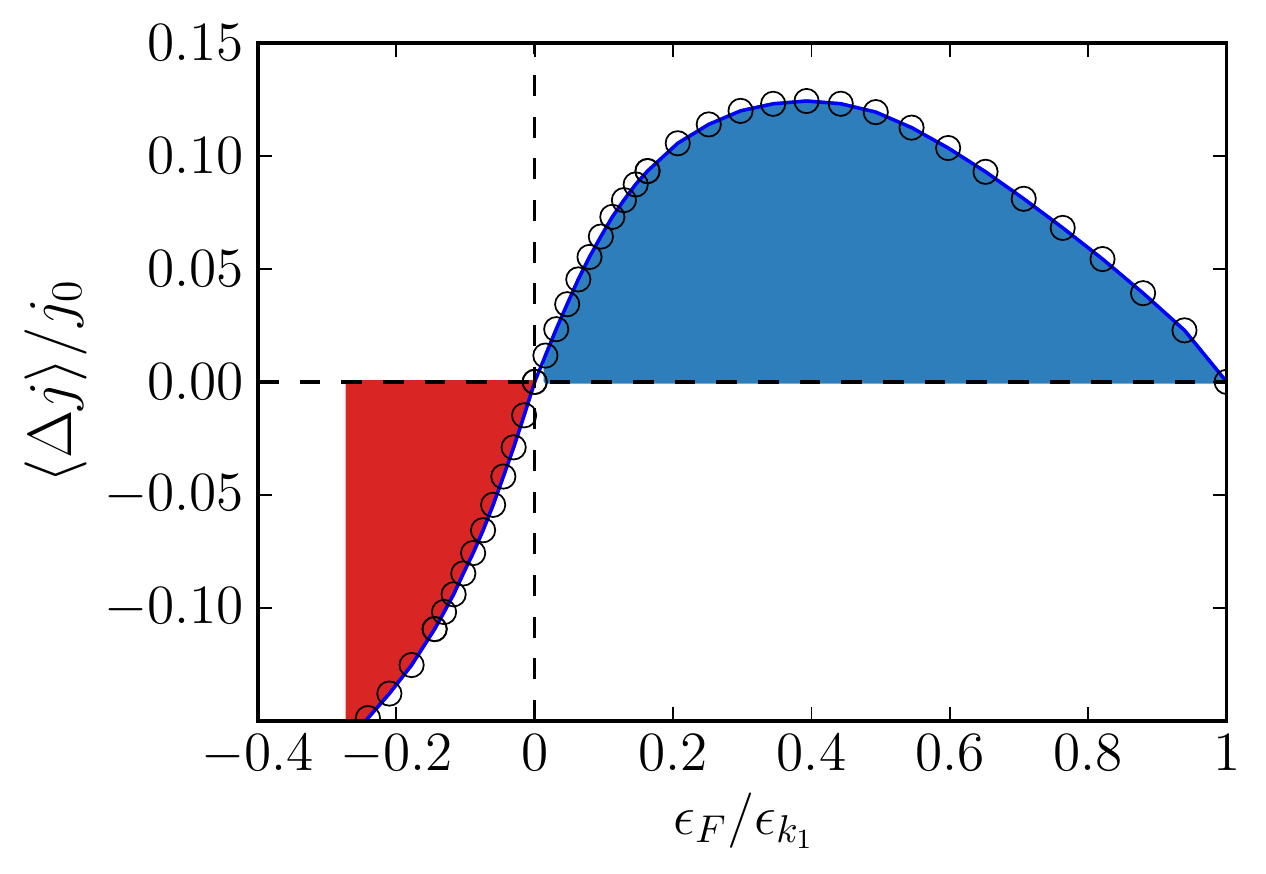}
\end{center}
\vspace{-3.45cm}
\hspace{3cm}
	\includegraphics[width=3.3cm]{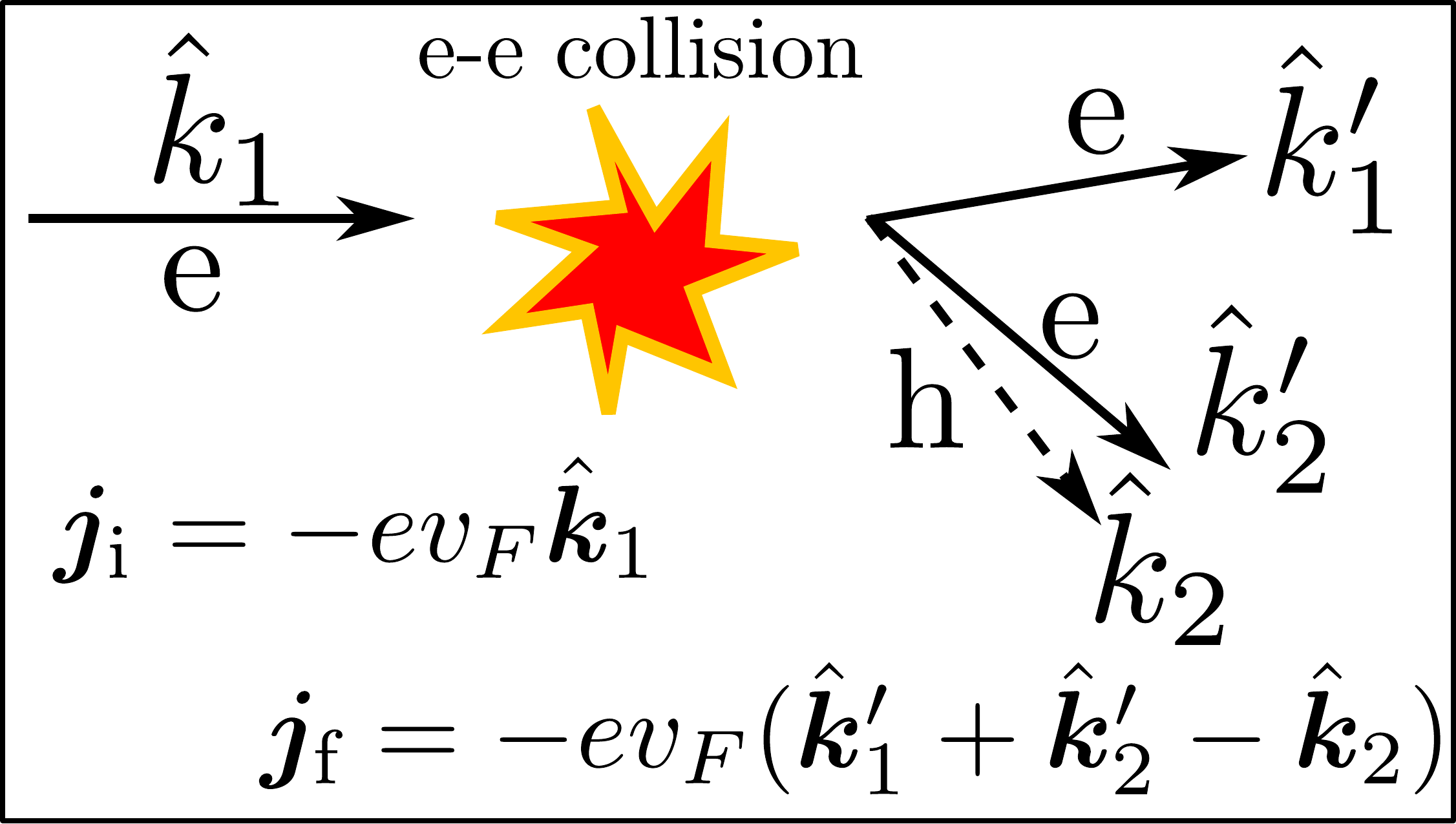}
\vspace{0.9cm}
\caption{(color online) Mean change of the current per electron scattering event $\langle\Delta j\rangle/j_0$ relative to the initial current $j_0$ {\em vs} $\epsilon_F$ for fixed initial energy $\epsilon_{\bm{k}_1} \approx0.15$ eV. The blue and red shaded areas indicate current amplification and relaxation, respectively. The rate of change of the current is normalized by the total scattering rate $\Gamma$. The results are obtained for realistic parameters for Bi$_2$Se$_3$, including particle-hole asymmetry $\xi=23.7$~eV~$\protect \text {\r {A}}^2$, $v_F=5\cdot 10^5$~m/s \cite{liu2010}, and $\alpha = 0.1$ \cite{greenaway1965,sandomirsky2001} (see text for definitions). Inset: Schematic of the excitation of an electron-hole pair by the initial photoexcited electron. }
\label{fig:meancurrkF}
\end{figure}

We find that this current relaxation process has rather surprising properties. Consider first the relaxation of the current associated with a single excited electron or hole above the Fermi sea. We find that the change in current strongly depends on the position of the Fermi level and, most importantly, changes sign as the Fermi energy crosses the Dirac point. This leads to the surprising conclusions that for an excited electron, the current actually increases rather than decreases by e-e collisions when the Fermi energy is above the Dirac point. Similarly, for an excited hole, the current increases when the Fermi energy lies below the Dirac point. The mean change of the current per scattering event is illustrated in Fig.~\ref{fig:meancurrkF} and is of the order of $\sim 10\%$ for realistic parameters. Ultimately, these remarkable results can be traced back to the fact that at zero temperature, there is no current relaxation when the Fermi energy is right at the Dirac point. 

For a photoexcited electron-hole pair, the relaxation of the total current involves a subtle interplay between the electron and the hole contribution. We find that overall, e-e scattering decreases the total current of the electron-hole pair, but the relaxation is strongly suppressed due to cancellations between electron and hole processes. In the limit of large excitation energies $\epsilon_1$ of the initial carriers, the rate of change of both electron and hole currents varies linearly with $\epsilon_F/\epsilon_1$, but with opposite signs. These linear terms cancel in the relaxation of the electron-hole pair and relaxation of the total current is dominated by subleading contributions which we find to scale as $\sim (\epsilon_F/\epsilon_1)^{3/2}$.

{\em Analysis of kinematic constraints.---}The surprising possibility of an increase in current due to e-e scattering can be seen most directly by  analyzing the kinematic constraints of the scattering process. Energy and momentum conservation demand
\begin{align}
\label{eq:econ0}
\epsilon_{\bm{k}_1}+\epsilon_{\bm{k}_2}&=\epsilon_{\bm{k}_1'}+\epsilon_{\bm{k}_2'},\\
 \bm{k}_1+\bm{k}_2&=\bm{k}_1'+\bm{k}_2'.
\label{eq:momconservation}
\end{align}
Here, $\bm{k}_i$ ($\bm{k}_i'$) is the momentum of the initial (final) electrons and $\epsilon_{\bm{k}} =\pm v_F k$ for the upper (conduction) and lower (valence) band, respectively. Expressed in terms of momentum, Eq.~\eqref{eq:econ0} depends on whether the specific scattering process is intraband or interband, with the allowed scattering processes depending on the Fermi energy.

If the Fermi energy lies at the Dirac point, $\epsilon_F=0$, a typical relaxation process of a highly excited electron is illustrated in Fig.~\ref{fig:relax1}(a). The excited electron relaxes by scattering off an electron in the Fermi sea, creating a hole in the valence band and an additional electron in the conduction band. Energy conservation, Eq.~\eqref{eq:econ0}, demands $k_1-k_2=k_1'+k_2'$, which takes into account that the electron in the Fermi sea has a negative energy, $\epsilon_{\bm{k}_2}=-v_F k_2$. Thus, the length of vector $\bm{k}_1$ must be equal to the sum of the lengths of the remaining three vectors. This is only satisfied for collinear scattering so that initial and final states have the same velocities, i.e., $\bm{v}_1=\bm{v}_2=\bm{v}_1'=\bm{v}_2'$, and the current remains unchanged in the e-e collision. For Dirac systems with the Fermi energy at the Dirac point, e-e interactions therefore do not relax current.

\begin{figure}[t]
 \includegraphics[width=5.5cm]{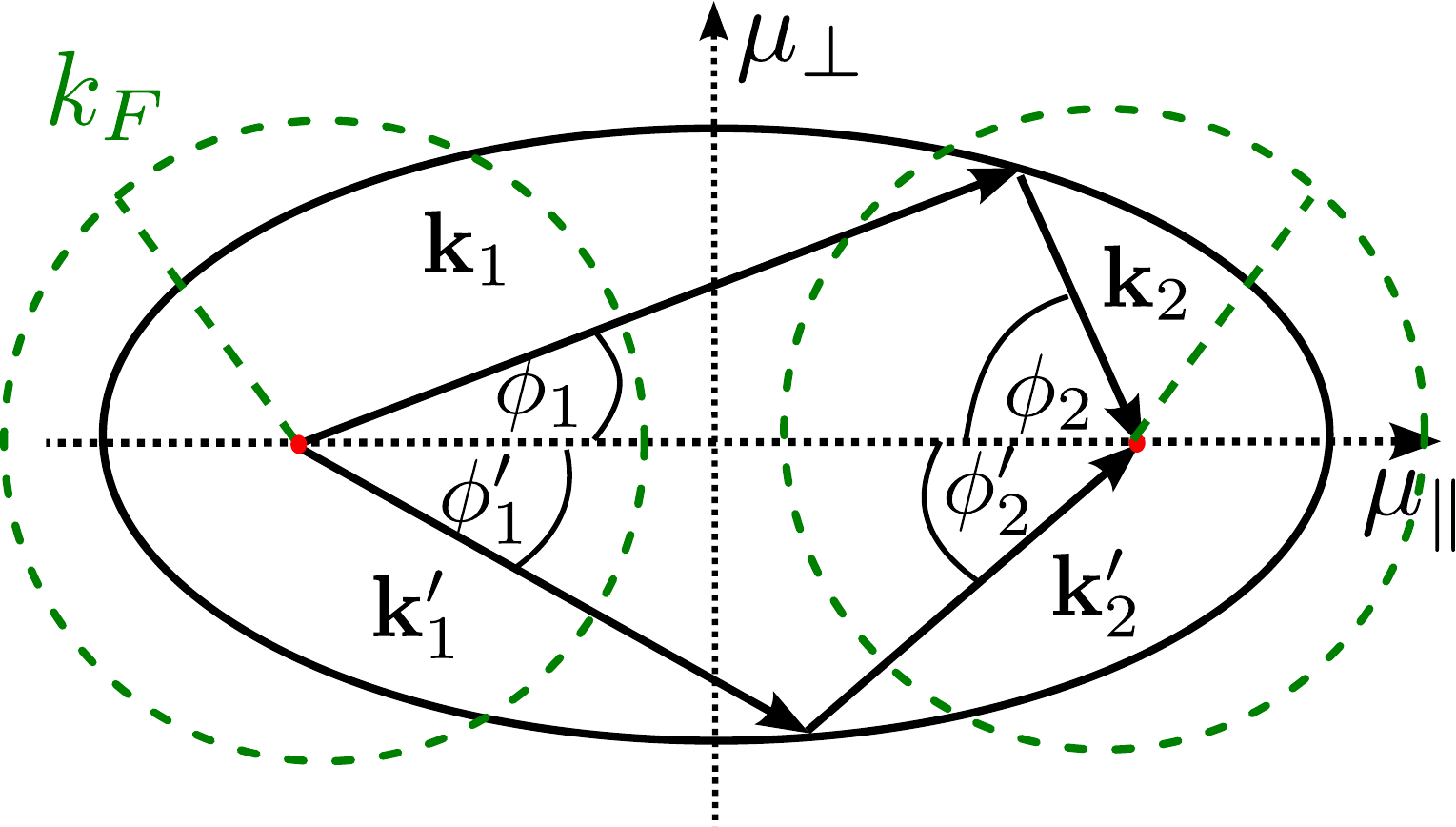}
 \caption{(color online) Kinematic ellipse for allowed electron scattering processes for $\epsilon_i>0$. $\bm{k}_1$ and $\bm{k}_2$ are drawn head-to-tail starting and ending at the left and right focal points (red dots). $\bm{k}_1'$ and $\bm{k}_2'$ are drawn in a similar fashion and touch at points that lie on the ellipse. $(k_1+k_2)/2$ and $\vert\bm{k}_1+\bm{k}_2\vert$ are the lengths of the semi-major axis and the distance between focal points, respectively. The green dashed circle indicates the Fermi momentum. Because of Pauli's principle we have $k_2\leq k_F<k_2',k_1'$ and the point where $\bm{k}_1'$ is connected to $\bm{k}_2'$ must lie on the ellipse outside the green dashed circles, while the point of connection of $\bm{k}_1$ and $\bm{k}_2$ has to lie inside the green dashed circle.}
 \label{fig:ellipse}
\end{figure}

When the Fermi energy lies above the Dirac point, i.e., $\epsilon_F>0$, the excited electron can interact with Fermi-sea electrons which are either in the conduction ($+$) or the valence ($-$) band [see Fig.~\ref{fig:relax1}(b)], corresponding to the processes $(+,+)\rightarrow(+,+)$ or $(+,-)\rightarrow(+,+)$. The previous argument for $\epsilon_F=0$ implies that a collision with electrons in the valence band, $(+,-)\rightarrow(+,+)$, is collinear and does not relax current. Thus, we only need to consider intraband processes [see Fig.~\ref{fig:relax1}(b)]. Energy conservation, i.e., $k_1+k_2=k_1'+k_2'$ from Eq.~\eqref{eq:econ0}, and momentum conservation \eqref{eq:momconservation} can now be graphically interpreted in terms of an ellipse as illustrated in Fig.~\ref{fig:ellipse}. The energy $k_1+k_2=k_1'+k_2'=const.$ defines the semi-major axis, while $\vert \bm{k}_1+\bm{k}_2 \vert=\vert \bm{k}_1'+\bm{k}_2' \vert$ is the distance between the focal points. 

This construction implies that the current actually increases along the direction of the initial current $\hatbm{k}_1$. In a first step, we assume that not only $\bm{k}_1$ but also $\bm{k}_2$ is fixed. The resulting ellipse is defined by the axes $\hat{\mu}_{\parallel}(\bm{k}_2)$ and $\hat{\mu}_{\perp}(\bm{k}_2)$ (see Fig.~\ref{fig:ellipse}). We will show that a summation over all $\bm{k}_1'$ and $\bm{k}_2'$ restricted to the ellipse leads to a current increase along $\hat{\mu}_{\parallel}$ as well as a change in current along $\hat{\mu}_{\perp}$. In a second step, we sum over all $\bm{k}_2$, i.e., over all ellipses. It turns out that a possible change of the current along $\hat{\mu}_{\perp}$ averages to zero due to the rotational symmetry of the problem. Remarkably, the increase in current along $\hat{\mu}_{\parallel}$ averages to an increase along the direction of the initial current $\hatbm{k}_1$ (see \cite{suppmat} for more details). 

The initial and final currents along $\hat{\mu}_{\parallel}$ are given by ($v_F=1$, $e=1$ for brevity)
\begin{eqnarray}
\label{eq:ji}
 j_{\textnormal{i}}&=&(\hatbm{k}_1)_{\mu_{\parallel}}=\cos\phi_1\\
 j_{\textnormal{f}}&=& (\hatbm{k}_1'+\hatbm{k}_2'-\hatbm{k}_2)_{\mu_{\parallel}}=\cos\phi_1'+\cos\phi_2'-\cos\phi_2,
\label{eq:jf}
\end{eqnarray}
where the angles are defined as in Fig.~\ref{fig:ellipse}. To analyze the change in current, we have to compare $\cos\phi_1+\cos\phi_2$ to $\cos\phi_1'+\cos\phi_2'$. It can be shown by elementary geometry that the sum of the cosines, restricted to the ellipse, has a maximum for the symmetric case where the of connection of the corresponding vectors lies on $\hat{\mu}_{\perp}$, and falls off monotonically away from the maximum. We know from Pauli's principle that $k_2\leq k_F<k_1',k_2'$. This implies that the point of connection of $\bm{k}_1$ and $\bm{k}_2$ must lie inside the green dashed circle of radius $k_F$ while the point of connection of $\bm{k}_1'$ and $\bm{k}_2'$ must lie outside this circle (see Fig.~\ref{fig:ellipse}) and thus closer to the $\hat{\mu}_{\perp}$-axis. Hence, $j_{\textnormal{f}} \geq j_{\textnormal{i}} $ for any scattering event, i.e., the current increases along $\hat{\mu}_{\parallel}$. Averaging over $\bm{k}_2$, i.e., averaging over all ellipses, leads to an average increase of the 
current along $\bm{k}_1$ \cite{suppmat}. An analogous argument shows that e-e scattering decreases the current when $\epsilon_F<0$ \cite{suppmat}.  

{\em Quantitative analysis.---}Quantitatively, the current relaxation rate for the optically excited electron-hole pair can be obtained within a golden-rule approach. For definiteness, we consider the surface states of the TI $\textnormal{Bi}_2\textnormal{Se}_3$, described by the (second-quantized) Dirac Hamiltonian
\begin{align}
 H&=\sum_{\bm{k}}\Psi^{\dagger}_{\bm{k}} \mathcal{H}_{\bm{k}} \Psi_{\bm{k}} +\frac{1}{2} \sum_{\bm{q},\bm{k}_1,\bm{k}_2} \Psi^{\dagger}_{\bm{k}_1+\bm{q}}\Psi^{\dagger}_{\bm{k}_2-\bm{q}} V(\bm{q})\Psi_{\bm{k}_2}\Psi_{\bm{k}_1},
\end{align}
where $V(q)=e^2/2 \varepsilon_0 \varepsilon q$ is the Coulomb interaction and 
\begin{equation}
\mathcal{H}_{\bm{k}}= v_F (k_x\sigma_y-k_y\sigma_x)
\label{eq:ham}
\end{equation}
describes the single-particle Dirac dispersion with eigenenergies $\epsilon_{\bm{k}}= \pm v_F k$ and eigenstates $\vert \bm{k}_i\rangle$. 

The initial photoexcitation creates an electron-hole pair with fixed momentum $\bm{k}_1$ [see Fig.\ \ref{fig:relax1}(a)]. Then, the rate of change of the electron and hole currents is
\begin{align}
 &\frac{d\bm{j}^{(e/h)}}{dt} = \mp e\sum_{\bm{k}_2,\bm{k}_1',\bm{k}_2'} (\bm{v}_1'+\bm{v}_2'-\bm{v}_1-\bm{v}_2) W_{\bm{k}_1,\bm{k}_2;\bm{k}_1',\bm{k}_2'}\nonumber\\
 &\,\,\,\,\,\,\times f^{(e/h)}(\epsilon_{\bm{k}_2})[1-f^{(e/h)}(\epsilon_{\bm{k}_1'})][1-f^{(e/h)}(\epsilon_{\bm{k}_2'})], 
 \label{eq:currderiv}
\end{align}
where the velocity is $\bm{v}_i=v_F \textnormal{sgn}(\epsilon_{\bm{k}_i}) \hatbm{k}_i$ and $f^{(e/h)}(\epsilon_{\bm{k}_i})$ denote the Fermi distribution function of electrons and holes, respectively. The transition rate is given by
\begin{align}
 W_{\bm{k}_1,\bm{k}_2;\bm{k}_1',\bm{k}_2'} =& \frac{2\pi}{\hbar}\vert M \vert^2 \delta_{\bm{k}_1+\bm{k}_2,\bm{k}_1'+\bm{k}_2'} \delta(\epsilon_1+\epsilon_2-\epsilon_1'-\epsilon_2'),
\label{eq:transrate}
\end{align}
with $\epsilon_i = \epsilon_{\bm{k}_i}$ and interaction matrix element
\begin{equation}
 M=\frac{1}{2L^2}[\langle \bm{k}_1'\vert \bm{k}_1\rangle\langle\bm{k}_2'\vert \bm{k}_2\rangle u(\vert \bm{k}_1-\bm{k}_1'\vert) 
 - (\bm{k}_1'\leftrightarrow\bm{k}_2')].
\end{equation}
Here, $L^2$ is the surface area of the system and $u(\bm{q})= (e^2/2 \varepsilon_0 \varepsilon)/( q+q_{\textnormal{TF}})$ the screened Coulomb interaction where $q_{\textnormal{TF}}=\alpha k_F$ is the Thomas-Fermi wave vector with $\alpha = e^2/(4\pi\hbar v_F \varepsilon_0 \varepsilon)$. As photoexcitation creates highly excited electron-hole pairs, we can set $T=0$.

Eq.~\eqref{eq:currderiv} can be simplified by introducing the momentum transfer $\bm{q}=\bm{k}_1-\bm{k}_1'=\bm{k}_2'-\bm{k}_2$ and the identity $\delta(\epsilon_1+\epsilon_2-\epsilon_1'-\epsilon_2')=\int d\omega \delta(\epsilon_1-\epsilon_1'-\omega)\delta(\epsilon_2'-\epsilon_2-\omega)$. Then, in the thermodynamic limit the two $\delta$-functions can be used to eliminate the angular integrals leaving us with a three-dimensional integral which can be solved numerically for general parameters and analytically in limiting cases \cite{suppmat}. 

We first evaluate the expressions numerically for a particular carrier type, namely the photoexcited electron, and compute the mean change of current per scattering event. To make our results realistic, we include the particle-hole asymmetry of the dispersion through $\epsilon_{\bm{k}}=\pm v_F k +\xi k^2$ \cite{note}. As illustrated in Fig.~\ref{fig:meancurrkF}, the current relaxes for negative Fermi energies but becomes amplified for positive Fermi energies, with the current 
enhancement being of the order of $\sim 10\%$.

\begin{figure}[t]
\centering
 \includegraphics[width=8.5cm]{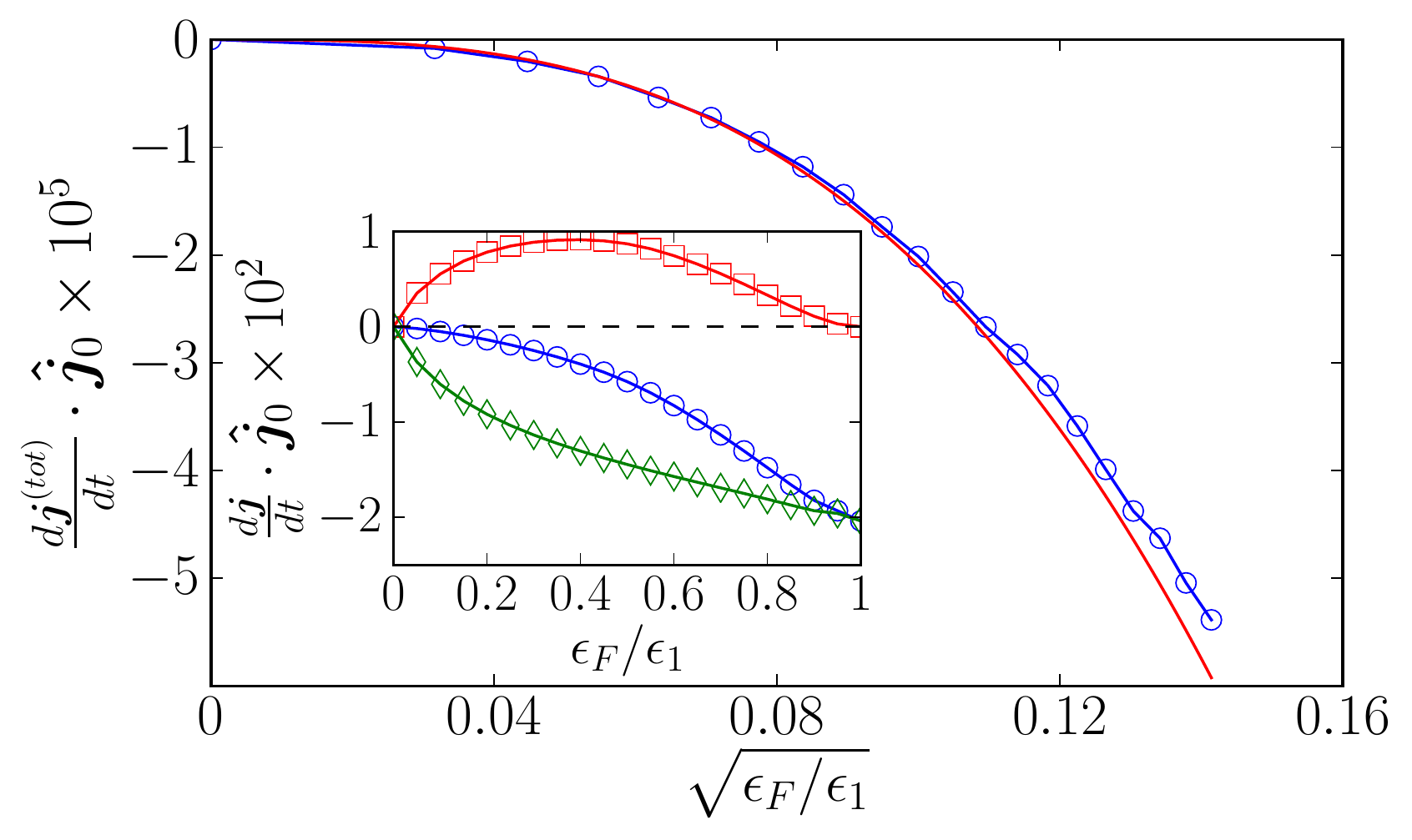}
\caption{(color online) Rate of change of the total current (blue circles) [from Eq.~\eqref{eq:currderiv}] and the asymptotic behavior $\sim \left(\epsilon_F/\epsilon_1\right)^{3/2}$ given by Eq.~\eqref{eq:currderivtot} (red). Inset: Rates of change of the electron (red squares), hole (green diamonds), and total (blue circles) currents. Parameters as in Fig.~\ref{fig:meancurrkF}. Relaxation of the total current is strongly suppressed due to cancellations between the electron and hole contributions. Results for $\epsilon_F<0$ follow by electron-hole symmetry.}
\label{fig:currderiv}
\end{figure}

From now on, we assume a perfectly linear dispersion, i.e., $\xi=0$, and $\epsilon_F>0$ for definiteness. The case of negative Fermi energy follows by electron-hole symmetry. The rates of change of the electron, hole, and total currents, obtained by numerically integrating Eq.~\eqref{eq:currderiv} are illustrated in the inset of Fig.~\ref{fig:currderiv}. As already seen, e-e scattering increases the electron current (red squares) and decreases the hole current (green diamonds). For large Fermi energies $\epsilon_F/\epsilon_1 \sim 1$ the rate of change approaches zero for the electron current and remains finite for the hole current, reflecting the different behavior of the phase space for scattering in the two cases [see Figs.~\ref{fig:relax1}(b) and (c)]. The e-e scattering also relaxes the total current (blue circles) but there are substantial cancellations between the electron and hole contributions. To quantify these cancellations, we analytically explore the asymptotic behavior of Eq.~\eqref{eq:currderiv}
for large excitation energies and small Fermi energy, i.e., for the limit $\epsilon_F/\epsilon_1 \ll 1$. Here, we focus on the results. Details of the calculations can be found in \cite{suppmat}.

For the electron current, we only need to consider the scattering process illustrated in Fig.~\ref{fig:relax1}(b) because interband processes, i.e., $(+,-)\rightarrow (+,+)$, are collinear and do not change the current, as shown above by the geometric argument. For the hole, we have to consider scattering processes like the one illustrated in Fig.~\ref{fig:relax1}(c). The hole can recombine with an electron in the valence band, exciting an electron from the conduction band above the Fermi energy, i.e., $(-,+)\rightarrow (-,+)$, or the hole can recombine with an electron in the conduction band, exciting an electron from the valence band above the Fermi energy, i.e., $(-,+)\rightarrow (+,-)$. Other allowed processes will be collinear. We find that the asymptotic behavior of the rate of change of the electron and hole currents is given by 
\begin{equation}
 \frac{d\bm{j}^{(e/h)}}{dt} \approx \pm C \alpha^2 \frac{\epsilon_F}{\hbar} \bm{j_0},
\label{eq:currderivlin}
\end{equation}
where $C\approx 0.3$ \cite{suppmat}, $\pm$ stands for the electron and hole current respectively, and $\bm{j_0}$ is the initial current of magnitude $j_0 = ev_F$ of the photoexcited carrier. This result has several interesting aspects. First, the time scale on which the initial current changes is independent of the large initial excitation energy $\epsilon_1$ of the photoexcited carrier and instead depends on the Fermi energy only. This is a consequence of the fact that the typical energy transfer in the relevant
e-e collisions is of the order of the Fermi energy. Secondly, to this order the rates of change of electron and hole currents differ only in their sign and thus cancel exactly. Thus, the rate of change of the total current of the photoexcited electron-hole pair is indeed much smaller and must scale with a higher power of $\epsilon_F/\epsilon_1$. We find that \cite{suppmat}
\begin{equation}
\frac{d\bm{j}^{(\textnormal{tot})}}{dt}  = \frac{d\bm{j}^{({e})}}{dt} + \frac{d\bm{j}^{({h})}}{dt} 
\approx- \frac{\alpha^2}{9} \frac{\epsilon_F}{\hbar} \left(\frac{\epsilon_F}{\epsilon_1}\right)^{1/2} \bm{j_0}.
\label{eq:currderivtot}
\end{equation}
The relaxation of the total current is suppressed for small $\epsilon_F/\epsilon_1$ and even vanishes in the limit $\epsilon_1 \rightarrow \infty$. Fig.~\ref{fig:currderiv} shows the rate of change of the total current for small $\epsilon_F/\epsilon_1$ determined by numerically integrating Eq.~\eqref{eq:currderiv} (blue circles), and the asymptotic behavior given by Eq.~\eqref{eq:currderivtot} (red solid line). 

These asymptotic behaviors of the total current and the individual electron and hole currents can be traced back to distinct scattering processes. The amplification and relaxation of the individual currents are governed by scattering processes with small energy transfers of the order of $\epsilon_F$. In contrast, for the total current the contributions with small energy transfer cancel exactly to the order considered and the result in Eq.~\eqref{eq:currderivtot} arises solely from scattering processes with large energy transfers of the order of $\epsilon_1$. Specifically, the relaxation of the total current is dominated by the interband hole process, $(-,+)\rightarrow (+,-)$, where the hole recombines with an electron from the conduction band while exciting an electron from the valence band to empty states in the conduction band. The predominance of scattering events with large energy transfers of the order of $\epsilon_1$ also explains why the relaxation vanishes in the 
limit of $\epsilon_1\rightarrow \infty$.

{\em Conclusion.---}Motivated by photocurrent measurements on various Dirac systems, we investigated the interaction-induced relaxation of photocurrents in clean Dirac systems and  uncovered a surprising effect: For a single excited electron, the current actually increases upon scattering with the electrons in the Fermi sea, as long as the chemical potential is above the Dirac point. Even for a single collision, this increase is substantial for realistic parameters. Furthermore, since high-energy electrons decay via a cascade of e-e collisions, the current increase can be further amplified by an additional factor up to $\sim \epsilon_1/\epsilon_F$, as the typical energy loss per collision is of the order of the Fermi energy \cite{song2013}. While the relaxation cascade underlying this argument has been predicted theoretically, the experimental situation remains inconclusive \cite{tani2012,gierz2013}, possibly due to competing optical-phonon collisions or a large radiation intensity which 
produces a high density of photoelectrons and phonons \cite{song2013,SongPrivate}.
Thus, the amplification effect may be most pronounced for excited electrons with energies below optical-phonon frequencies ($\approx200$~meV for graphene) and for low-intensity irradiation. This current amplification has important implications for photocurrents where it results in a substantial suppression of the current-relaxation rate of photoexcited electron-hole pairs, but might also be observable more directly in other types of experiments. Most promising may be time-resolved measurements of photocurrents as recently performed on graphene \cite{sun2012}. One might also expect a strong non-linear signature in IV characteristics, since high energy electrons produce a jet of induced current. For the same reason, the current amplification might enhance the photoconductivity (electron conductivity in the presence of light), with the effect increasing with the frequency of the irradiating light. We intend to pursue a quantitative analysis of such effects in future work. 

{\em Acknowledgements.---}We thank J.\ Eisenstein, Erik Henriksen, Justin Song and Feng Wang for discussions and acknowledge financial support through SPP 1666 of the Deutsche Forschungssemester and a Helmholtz Virtual Institute ``New States of Matter and Their Excitations'' (Berlin) as well as DARPA, the IQIM, an NSF institute supported by the Moore Foundation, and the Humboldt Foundation (Pasadena). 

\bibliographystyle{apsrev}

\onecolumngrid
\newpage
\begin{center}
 \textbf{Current amplification and relaxation in Dirac systems - Supplemental Material}
\end{center}

\section{Table of contents}
\begin{enumerate}[A)]
 \item Details of the geometric argument for the current increase or decrease
 \item Evaluation of the energy conservation $\delta$-function
 \item Identification of distinct scattering processes in the asymptotic behavior of the rate of change of the current
 \item Asymptotic behavior of the rate of change of the current for large excitation energies
\item Definition of the mean change of the current per electron scattering event
\end{enumerate}

\section{A) Details of the geometric argument for current increase or decrease}

\begin{figure}[b]
 \includegraphics[width=15cm]{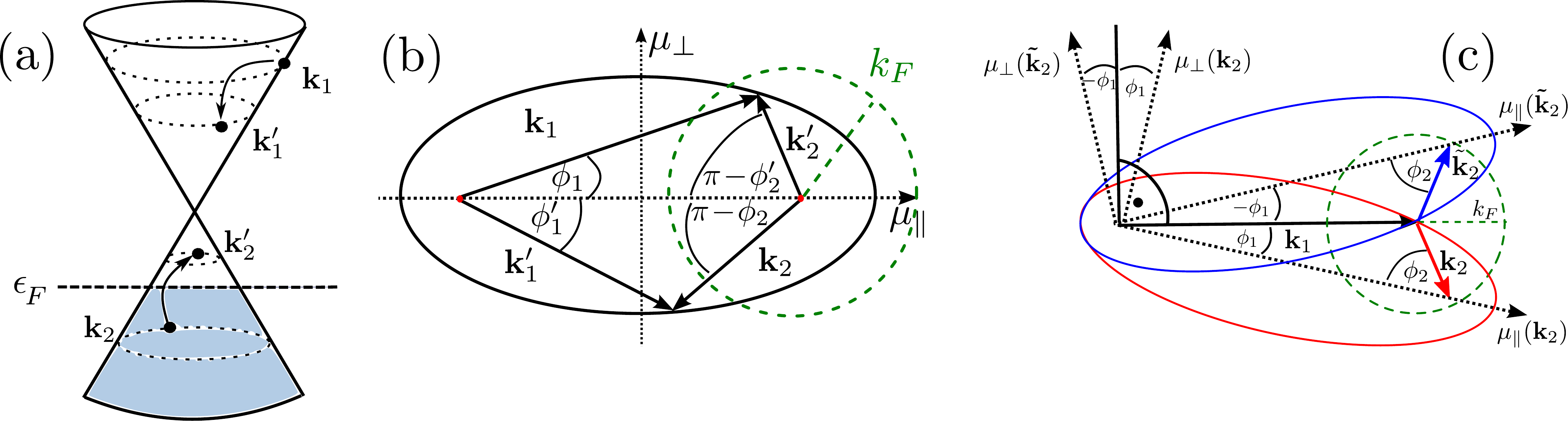}
 \caption{(color online) a) Schematic illustration of two-electron scattering processes where a hot
electron relaxes by creating an electron-hole pair for $\epsilon_F<0$. The dashed circles indicate constant energy contours and are a guide to the eye. Processes where the final states $\bm{k}_1'$ and $\bm{k}_2'$ are both either in the lower or upper band are similar to the case of $\epsilon=0$ and do not affect the current. b) For $\epsilon_F<0$ the allowed scattering processes can also be represented by an ellipse. $(k_1+k_2')/2$ is the length of the semi-major axis and $\vert \bm{k}_1-\bm{k}_2'\vert$ is the distance between the focal points. The green dashed circle again indicates the Fermi momentum. Because of Pauli's principle we have $k_2'< k_F \leq k_2$. Note that here we do not have a restriction on $k_1'$ other than $k_1'\leq k_1$. c) When averaging over all allowed $\bm{k}_2$, each $\bm{k}_2$ parametrizes a different ellipse. For each $\bm{k}_2$ and the resulting ellipse (red) there is a mirror image
with respect to $\bm{k}_1$, $\tilde{\bm{k}}_2$ (blue), such that an increase in the current along $\hat{\mu}_{\parallel}(\bm{k}_2)$ averages to an increase along $\hatbm{k}_1$. In general there can also be a change of the current in direction $\hat{\mu}_{\perp}$. The change in the component parallel to $\bm{k}_1$, however, averages to zero when averaging over $\bm{k}_2$. By symmetry there can be no change in current perpendicular to $\bm{k}_1$. Changes of the current in the $\hat{\mu}_{\perp}$ direction are therefore not important for the average change in current. The green dashes circle indicates the Fermi momentum. The allowed states $\bm{k}_2$ must lie within this circle.}
 \label{fig:suppmatellipse}
\end{figure}

For the case of $\epsilon_F>0$ we have shown that for an arbitrary $\bm{k}_2$ the component of the current along the direction given by the major axis of that specific ellipse, i.e., along $\hat{\mu}_{\parallel}(\bm{k}_2)$, increases due to e-e scattering. In general, the component of the current along $\hat{\mu}_{\perp}(\bm{k}_2)$ might also change during a scattering event. However, we still need to sum over all $\bm{k}_2$. Summing over $\bm{k}_2$ means summing over all possible ellipses and thus over all possible $\hat{\mu}_{\parallel}$ and $\hat{\mu}_{\perp}$. 
For a given $\bm{k}_2$ and resulting $\phi_1$, by symmetry there is also a $\bm{\tilde{k}}_2$, i.e., the mirror image of $\bm{k}_2$ with respect to an axis parallel to $\bm{k}_1$, that leads to $-\phi_1$ as shown in Fig.~\ref{fig:suppmatellipse}(c). Thus, when summing over all possible $\bm{k}_2$, the increase of the component of the current along $\hat{\mu}_{\parallel}(\bm{k}_2)$ averages to a current increase along $\hatbm{k}_1$. A change in the component along $\hat{\mu}_{\perp}(\bm{k}_2)$ has components parallel and perpendicular to $\bm{k}_1$. The component parallel to $\bm{k}_1$ changes sign under the described reflection and thus averages to zero. The component perpendicular to $\bm{k}_
1$ also has to average to zero because the rotational symmetry of our systems requires that the average change in current can only be in the $\hatbm{k}_1$-direction. By symmetry there can be no change in current perpendicular to $\hatbm{k}_1$.

An analogous argument to the case of $\epsilon_F>0$ can be made for the case of $\epsilon_F<0$. Here the highly excited electron in the upper band scatters off an electron in the Fermi sea in the lower band. The scattering process that affects the current has final electronic states which are in different bands as illustrated in Fig.~\ref{fig:suppmatellipse}(a), i.e., $(+,-)\rightarrow (+,-)$. Processes where both final states are in the same band are equivalent to the process for $\epsilon_F=0$ and thus do not change the current. The condition for energy conservation becomes $k_1-k_2=k_1'-k_2'$. Written in the following way, energy and momentum conservation can again be represented by an ellipse as illustrated in Fig.~\ref{fig:suppmatellipse}(b),
\begin{align}
k_1+k_2'&=k_1'+k_2 \\
\bm{k}_1-\bm{k}_2'&=\bm{k}_1'-\bm{k}_2.
\end{align}

Note that in Fig.~\ref{fig:suppmatellipse}(b) the orientations of $\bm{k}_2$ and $\bm{k}_2'$ are reversed with respect to the analogous case of $\epsilon_F>0$, and we have $k_2'<k_F \leq k_2$ (see Fig.~\ref{fig:suppmatellipse}(a)). The initial and finals currents along $\hat{\mu}(\bm{k}_2')$ (see Fig.~\ref{fig:suppmatellipse}(b)) are thus given by $j_{\textnormal{i}}=\cos\phi_1$ and $j_{\textnormal{f}}=\cos\phi_1'-\cos\phi_2'+\cos\phi_2$ and we have to compare $\cos\phi_1'+\cos\phi_2$ to $\cos\phi_1+\cos\phi_2'$ to analyze the change in current. Because $k_2'<k_F \leq k_2$ due to Pauli's principle, the point of connection of $\bm{k}_1$ and $\bm{k}_2'$, as drawn in Fig.~\ref{fig:suppmatellipse}(b), will be inside the green dashed circle while the point of connection of $\bm{k}_1'$ and $\bm{k}_2$ will lie outside this circle. Again it can be shown by elementary geometry that this implies $j_{\textnormal{f}}-j_{\textnormal{i}} \leq 0$ for any scattering event, i.e., the current decreases along $\hat{\mu}_{\
parallel}$. 
Analogously to above, averaging over all possibles scattering processes leads to a current decrease due to e-e scattering for $\epsilon_F<0$.


\section{B) Evaluation of the energy conservation $\delta$-function}

After introducing the momentum transfer $\bm{q}=\bm{k}_1-\bm{k}_1'=\bm{k}_2'-\bm{k}_2$ the energy conservation $\delta$-function can be written as
\begin{align}
\delta(\epsilon_{\bm{k}_1}+\epsilon_{\bm{k}_2}-\epsilon_{\bm{k}_1-\bm{q}}-\epsilon_{\bm{k}_2+\bm{q}})&=\frac{1}{\hbar v_F}\int dp \delta(k_1-\absa-p)\delta(\absb-k_2-p),
\end{align}
where $\hbar v_F p$ is the difference in energy of the initial and final scattering states.
Using the relation $\delta(a-b)=2a\delta(a^2-b^2)$ with $a,b>0$, we can write
\begin{align}
\delta(k_1-\vert \bm{k}_1-\bm{q}\vert -p)&=2 (k_1-p)  \delta((k_1-p)^2-\absa^2)\nonumber\\
&= 2 (k_1-p) \int_0^{2\pi} d\phi_q f(\cos\phi_q) \delta(-2 k_1p +p^2-q^2+2 k_1 q \cos\phi_q)\nonumber\\
&=\frac{(k_1-p)}{k_1 q}  \delta(\cos\phi_q-\frac{q^2-p^2 +2 k_1p}{2 k_1 q})\theta\left(1-\left\vert \frac{q^2-p^2 +2 k_1p}{2 k_1 q} \right\vert \right)
\label{eq:delta1}
\end{align}
where $\phi_q$ is the angle between $\bm{k}_1$ and $\bm{q}$, such that

\begin{align}
\int_0^{2\pi} d\phi_q f(\cos\phi_q)\delta(k_1-\vert \bm{k}_1-\bm{q}\vert -p)&=2 \frac{(k_1-p)}{k_1 q} \frac{1}{\sqrt{1-(\frac{q^2-p^2 +2 k_1p}{2 k_1 q})^2}}f\left( \frac{q^2-p^2 +2 k_1p}{2 k_1 q} \right)\theta\left(1-\left\vert \frac{q^2-p^2 +2 k_1p}{2 k_1 q} \right\vert \right)
\label{eq:delta2}
\end{align}
where the factor of $2$ comes from the fact that $\cos\phi -a$ has two zeroes in the interval $[0,2\pi]$ with $\vert a \vert \leq 1$. Analogously, the $\phi_2$-integration can be performed evaluating $\delta(\absb-k_2-p)$.

\section{C) Identification of distinct scattering processes in the asymptotic behavior of the rate of change of the current}

\begin{figure}[ht]
\centering
 \includegraphics[width=7cm]{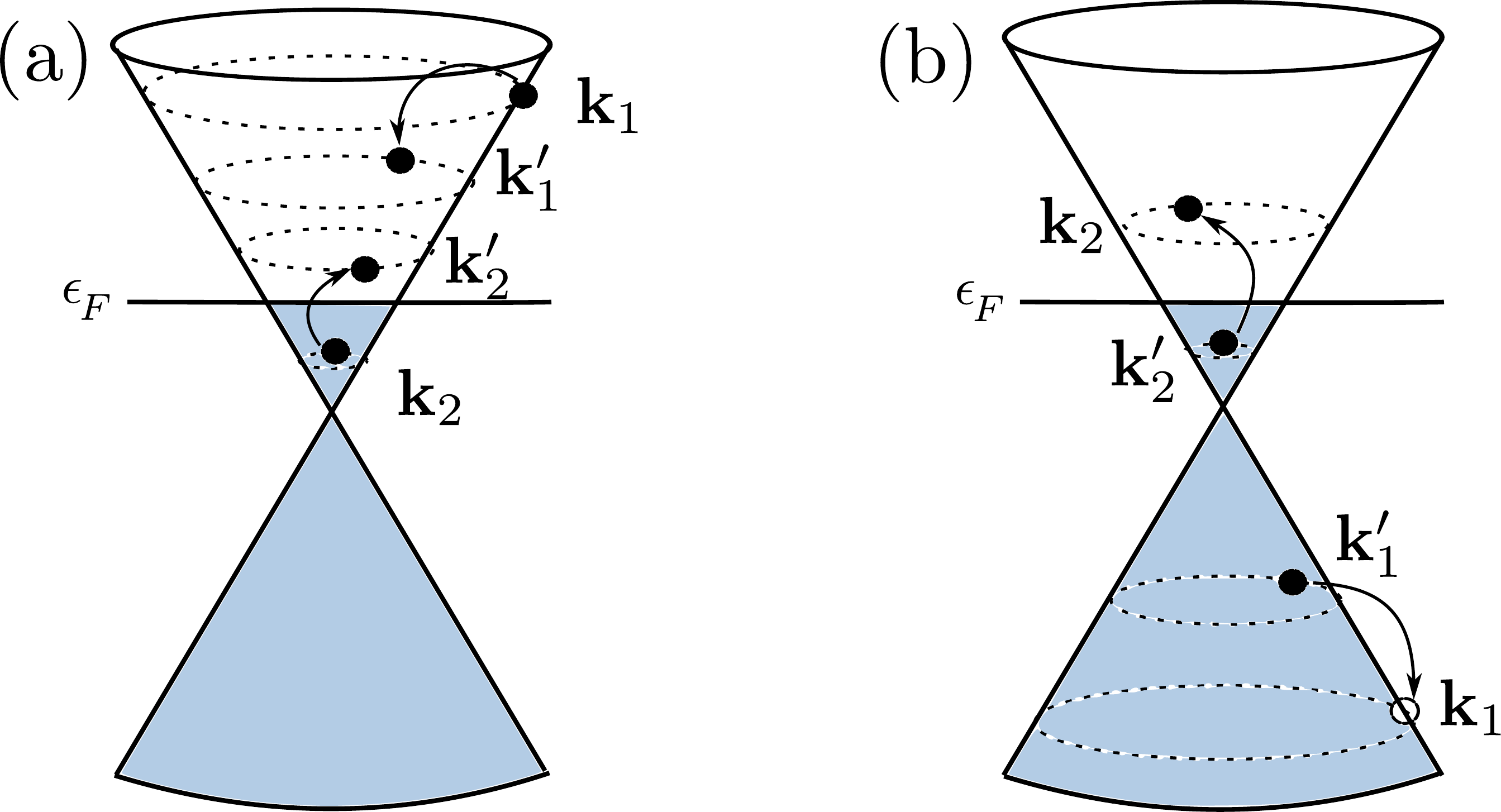}
 \caption{a) Electron and b) hole process for the case of $\epsilon_F>0$. a) Scattering processes where $\vert \bm{k}_2\rangle$ is in the lower band are collinear and will not change the current. b) Processes where $\vert \bm{k}_1'\rangle$, $\vert \bm{k}_2'\rangle$ are in the same band are also collinear.}
 \label{fig:holeprocess}
\end{figure}

For $\epsilon_F>0$ we only need to consider processes like the one illustrated in Fig.~\ref{fig:holeprocess}(a), in which the excited electron scatters off an electron in the upper band. Other allowed processes will be collinear and thus do not change the current. The rate of change of the electron current is then given by
\begin{align}
 \frac{d \bm{j}^{e}}{dt}&= -e \frac{1}{4 L^4}\frac{2 \pi}{\hbar} \left(\frac{e^2}{2 \epsilon_0\epsilon}\right)^2\sum_{\bm{k}_2,\bm{k}_2',\bm{k}_1'}\delta(\epsilon_{\bm{k}_1}+\epsilon_{\bm{k}_2}-\epsilon_{\bm{k}_1'}-\epsilon_{\bm{k}_2'}) \left(\hat{k}_1'+\hat{k}_2'-\hat{k}_1-\hat{k}_2\right) \nonumber\\
&\quad \times \left\vert \frac{\langle \bm{k}_1,+\vert\bm{k}_1',+\rangle\langle \bm{k}_2,+\vert\bm{k}_2',+\rangle}{\vert \bm{k}_1-\bm{k}_1'\vert+q_{TF}} - \frac{\langle \bm{k}_1,+\vert\bm{k}_2',+\rangle\langle \bm{k}_2,+\vert\bm{k}_1',+\rangle}{\vert \bm{k}_1-\bm{k}_2'\vert+q_{TF}}\right\vert^2 \theta(\epsilon_F-\epsilon_{\bm{k}_2})\theta(\epsilon_{\bm{k}_2'}-\epsilon_F)\theta(\epsilon_{\bm{k}_1'}-\epsilon_F) \nonumber\\
&=-e \frac{1}{4 L^4}\frac{2 \pi}{\hbar} \left(\frac{e^2}{2 \epsilon_0\epsilon}\right)^2\sum_{\bm{k}_2,\bm{k}_2',\bm{k}_1'}\delta(\epsilon_{\bm{k}_1}+\epsilon_{\bm{k}_2}-\epsilon_{\bm{k}_1'}-\epsilon_{\bm{k}_2'}) \left(\hat{k}_1'+\hat{k}_2'-\hat{k}_1-\hat{k}_2\right) \left(\vert M^e_d- M^e_{ex}\vert^2\right) \nonumber\\
&\quad \times \theta(\epsilon_F-\epsilon_{\bm{k}_2})\theta(\epsilon_{\bm{k}_2'}-\epsilon_F)\theta(\epsilon_{\bm{k}_1'}-\epsilon_F),
\label{eq:ederiv1}
\end{align}
where the sum is over states with positive energy only and we decide to call the first term in the interaction matrix element 'direct' and the second 'exchange'. Performing the sum, we find that the contributions to the rate of change from $\vert M_d^e\vert^2$ and $\vert M_{ex}^e\vert^2$ are equal, as can be easily seen by switching the labels $\bm{k}_1' \leftrightarrow \bm{k}_2'$ in one of the terms. We will call these contributions $d\bm{j}^e_d/dt$ and $d\bm{j}^e_{ex}/dt$ respectively, with $d\bm{j}^e_d/dt=d\bm{j}^e_{ex}/dt$. The remaining contribution from the interference term proportional to $2 Re[M_d^e(M_{ex}^e)^*]$ we will call $d\bm{j}^e_{\textnormal{inter}}/dt$.


Analogously for the hole current, we only need to consider processes where the hole recombines with an electron in the lower(upper) band thereby exciting an electron from the upper(lower) band above the Fermi energy, i.e., the states $\vert \bm{k}_1'\rangle$, $\vert\bm{k}_2'\rangle$ of the scattering event are in different bands as illustrated in Fig.~\ref{fig:holeprocess}(b). Processes where $\vert \bm{k}_1'\rangle$, $\vert\bm{k}_2'\rangle$ are in the same band are collinear and thus do not change the current. The rate of change of the hole current can be written as
\begin{align}
\frac{d \bm{j}^{h}}{dt}&= 2 e \frac{1}{4 L^4}\frac{2 \pi}{\hbar} \left(\frac{e^2}{2 \epsilon_0\epsilon}\right)^2\sum_{\bm{k}_2,\bm{k}_2',\bm{k}_1'}\delta(\epsilon_{\bm{k}_1}+\epsilon_{\bm{k}_2}-\epsilon_{\bm{k}_1'}-\epsilon_{\bm{k}_2'}) \left(-\hat{k}_1'+\hat{k}_2'+\hat{k}_1-\hat{k}_2\right) \nonumber\\
&\quad\times \left\vert \frac{\langle \bm{k}_1,-\vert\bm{k}_1',-\rangle\langle \bm{k}_2,+\vert\bm{k}_2',+\rangle}{\vert \bm{k}_1-\bm{k}_1'\vert+q_{TF}} - \frac{\langle \bm{k}_1,-\vert\bm{k}_2',+\rangle\langle \bm{k}_2,+\vert\bm{k}_1',-\rangle}{\vert \bm{k}_1-\bm{k}_2'\vert+q_{TF}}\right\vert^2 \theta(\epsilon_{\bm{k}_2}-\epsilon_F)\theta(\epsilon_F-\epsilon_{\bm{k}_2'}) \nonumber\\
&=2 e \frac{1}{4 L^4}\frac{2 \pi}{\hbar} \left(\frac{e^2}{2 \epsilon_0\epsilon}\right)^2\sum_{\bm{k}_2,\bm{k}_2',\bm{k}_1'} \delta(\epsilon_{\bm{k}_1}+\epsilon_{\bm{k}_2}-\epsilon_{\bm{k}_1'}-\epsilon_{\bm{k}_2'}) \left(-\hat{k}_1'+\hat{k}_2'+\hat{k}_1-\hat{k}_2\right) \nonumber\\
&\quad \times \left( \vert M^h_{d} - M^h_{ex}\vert^2 \right) \theta(\epsilon_{\bm{k}_2}-\epsilon_F)\theta(\epsilon_F-\epsilon_{\bm{k}_2'}),
\label{eq:hderiv1}
\end{align}
where we restricted the sum to $\epsilon_{\bm{k}_1'}<0$ and $\epsilon_{\bm{k}_2'}>0$ and added the factor of $2$ in front for the other half of the sum. We again call the first term of the interaction matrix element 'direct' and the second 'exchange'. Analogously to above we call the corresponding contributions to the rate of change of the hole current $d\bm{j}^h_d/dt$ and $d\bm{j}^h_{ex}/dt$, and $d\bm{j}^h_{\textnormal{inter}}/dt$. Here, the contributions from direct and exchange term are not equal because the states $\vert \bm{k}_1'\rangle$, $\vert\bm{k}_2'\rangle$ are in different bands. Switching the labels as for the electron current does not transform one term into the other.

For large excitation energies, i.e., for $\epsilon_F/\epsilon_1 \ll 1$, we find to lowest order that 
\begin{eqnarray}
 \frac{d\bm{j}_d^e}{dt}+\frac{d\bm{j}_{ex}^e}{dt} &\approx&-\frac{d\bm{j}_d^h}{dt}, \\
 \frac{d\bm{j}_{\textnormal{inter}}^e}{dt} &\approx& -\frac{d\bm{j}_{\textnormal{inter}}^e}{dt}. 
\end{eqnarray}
As shown in the next section this cancellation results in the fact that the rate of change of the total current to leading order is simply given by 
\begin{equation}
  \frac{d\bm{j}^{\textnormal{tot}}}{dt}=\frac{d\bm{j}^{e}}{dt}+\frac{d\bm{j}^{h}}{dt} \approx \frac{d\bm{j}_{ex}^h}{dt}.
\end{equation}

$d\bm{j}^h_{ex}/dt$ is governed by the interaction matrix element
\begin{equation}
 \vert M^h_{ex}\vert^2=\left\vert \frac{\langle \bm{k}_1,-\vert\bm{k}_2',+\rangle\langle \bm{k}_2,+\vert\bm{k}_1',-\rangle}{\vert \bm{k}_1-\bm{k}_2'\vert+q_{TF}}\right\vert^2,
\end{equation}
which describes processes where the photoexcited hole recombines with an electron from the upper band thereby exciting an electron from the lower band above the Fermi energy. These scattering processes involve large energy transfers of the order of the initial excitation energy $\epsilon_1$.


\section{D) Asymptotic behavior of the rate of change of the current for large excitation energies}
We now want to calculate the asymptotic behavior of the rates of change of the individual electron and hole currents and of the total current. We will show the calculation for $d\bm{j}^e_d/dt$ in detail. The calculations of the remaining contributions follow analogously. 

After introducing the momentum transfer $\bm{q}=\bm{k}_1-\bm{k}_1'=\bm{k}_2'-\bm{k}_2$, we use that the Coulomb interaction in the direct term of Eq.~\eqref{eq:ederiv1} is proportional to $\sim 1/(q+q_{TF})$ and the integral will be dominated by scattering events with small momentum transfer $q \ll k_1$. We now fix the initial momentum of the excited electron-hole pair $\bm{k}_1=k_1\hat{x}$ such that the initial current is given by $\bm{j}_0=-2ev \hat{x}$. The difference of the velocities of states $\bm{k}_1$ and $\bm{k}_1'$ can be approximated by zero, i.e.,  
\begin{equation}
\hat{k}_1'-\hat{k}_1=\frac{k_1-q\cos\phi_q}{\vert \bm{k}_1-\bm{q}\vert }-1 \approx 0.
\label{eq:approx1}
\end{equation}
We can also approximate the spin overlap of states $\bm{k}_1$ and $\bm{k}_1'$ by $1$, i.e.,
\begin{equation}
\vert \langle \bm{k}_1 \vert\bm{k}_1'\rangle \vert^2=\frac{k_1-q\cos\phi_q+\absa}{2\absa} \approx 1.
\label{eq:approx2}
\end{equation}
The initial current flows in the negative $x$-direction so the rate of change of the current will only have an $x$-component. By the rotational symmetry of our problem there can be no change of the current along $\hat{y}$. The sum of the quadratic direct and exchange contributions to the rate of change of the electron current given by Eq.~\eqref{eq:ederiv1} can then be written as 
\begin{align}
 \frac{d j^{e}_d}{dt}+\frac{d j^{e}_{ex}}{dt}&= -e v \frac{1}{2 L^4}\frac{2 \pi}{\hbar} \left(\frac{e^2}{2 \epsilon_0\epsilon}\right)^2\sum_{\bm{k}_2,\bm{q}}\frac{1}{\hbar v}\delta(k_1+k_2-\vert \bm{k}_1-\bm{q}\vert -\vert \bm{k}_2+\bm{q}\vert) \left(\frac{k_2\cos(\phi_2+\phi_q)+q\cos\phi_q}{\absb}-\cos(\phi_2+\phi_q) \right) \nonumber\\
&\quad \times \frac{1}{(q+\alpha k_F)^2} \left(\frac{k_2+q\cos\phi_2+\absb}{2\absb} \right) \theta(k_F-k_2)\theta(\vert \bm{k}_2+\bm{q}\vert-k_F),
\label{eq:ederivlin2}
\end{align}
where $\alpha=e^2 /(4 \pi \hbar v_F \varepsilon_0\varepsilon)$ is defined by $q_{\textnormal{TF}}= \alpha k_F$.
As above, we use the identity
\begin{equation}
\delta(k_1+k_2-\vert \bm{k}_1-\bm{q}\vert -\vert \bm{k}_2+\bm{q}\vert) =\int dp \delta(k_1-\absa -p)\delta(\absb-k_2-p)
\end{equation}
and evaluate the $\phi_2$ and $\phi_q$ integrations with the two $\delta$-functions as shown in Eqs.~\eqref{eq:delta1} and \eqref{eq:delta2}. In Eq.~\eqref{eq:ederivlin2}, however, we not only have terms with $\cos\phi_2$ and $\cos\phi_q$ but also terms that contain $\sin\phi_2 \sin\phi_q$. Depending on the values of $\phi_2$ and $\phi_q$, we can write
\begin{equation}
\sin\phi_2 \sin\phi_q=\pm \sqrt{1-\cos^2\phi_2}\sqrt{1-\cos^2\phi_q}.
\end{equation}

Since we have to integrate both $\phi_2$ and $\phi_q$ from $0$ to $2\pi$, integration of the terms proportional to $\sin\phi_2\sin\phi_q$ gives zero. Thus, in the integrand of Eq.~\eqref{eq:ederivlin2} we can neglect the terms proportional to $\sin\phi_2\sin\phi_q$, leaving us with a function that only depends on $\cos\phi_2$ and $\cos\phi_q$. Performing the $\phi_2$ and $\phi_q$ integrations using Eq.~\eqref{eq:delta2} and simplifying the result, we are left with 

\begin{align}
 \frac{d j^{e}_d}{dt}+\frac{d j^{e}_{ex}}{dt}&=-e v \frac{1}{2}\frac{2 \pi}{\hbar} \left(\frac{e^2}{2 \epsilon_0\epsilon}\right)^2\frac{1}{\hbar v}\frac{1}{(2\pi)^4} \left(\int_{0}^{k_F} dp \int_{k_F-p}^{k_F} dk_2 + \int_{k_F}^{k_1-k_F} dp\int_{0}^{k_F} dk_2 \right)\int_{p}^{2k_2+p} dq \nonumber\\
&\quad \times  (k_1-p)\frac{\sqrt{(2k_2+p)^2-q^2}}{\sqrt{(2k_1-p)^2-q^2}} \left(\frac{q^2-p^2+2k_1p}{k_1q}\right)\frac{1}{(q+\alpha k_F)^2} \frac{(2k_2+p)}{k_2(k_2+p)}.
\label{eq:ederivlin5}
\end{align}

For the integral over small $p \ll k_1$ we can approximate the integrand further. Introducing dimensionless parameters $\bar{p}=p/k_F$, $\bar{q}=q/k_F$, and $\bar{k_2}=k_2/k_F$ and shifting the integration variable $k_2\rightarrow k_2+p/2$, we get


\begin{equation}
-\frac{D}{4\pi} \frac{\epsilon_F}{\epsilon_1} \int_{0}^{1} d\bar{p} \int_{1-\bar{p}/2}^{1+\bar{p}/2} d\bar{k_2} \int_{\bar{p}}^{2\bar{k_2}} d\bar{q}  \sqrt{(2\bar{k_2})^2-\bar{q}^2}\left(\frac{\bar{p}}{\bar{q}}\right)\frac{1}{(\bar{q}+\alpha)^2} \frac{2\bar{k_2}}{\bar{k_2}^2-\frac{\bar{p}^2}{4}}\approx -\gamma \frac{D}{4\pi} \frac{\epsilon_F}{\epsilon_1}.
\label{eq:ederivlinsmallomega}
\end{equation}
with $\gamma \approx 1.17$ from numerical integration and $D=e v_F\alpha^2 \epsilon_1/\hbar$ and $\alpha=0.1$ as in the main text.

For the remaining integral we also use the dimensionless parameters $\bar{p}$, $\bar{q}$, and $\bar{k_2}$. If the integral converges, then the integrand has to go to zero faster than $1/p$, i.e., the weight is negligible for large $p$ and we are still allowed to approximate $p \ll k_1$. The integral becomes

\begin{align}
-\frac{D}{4\pi} \frac{\epsilon_F}{\epsilon_1}  \int_{1}^{k_1/k_F-1} dp\int_{0}^{1} d\bar{k_2}\int_{\bar{p}}^{2\bar{k_2}+\bar{p}} dq  \sqrt{2p} \sqrt{2\bar{k_2}+\bar{p}-\bar{q}}\left(\frac{\bar{p}}{\bar{q}}\right)\frac{1}{(\bar{q}+\alpha)^2} \frac{(2\bar{k_2}+\bar{p})}{\bar{k_2}(\bar{k_2}+\bar{p})}.
\label{eq:ederivlinlargeromega}
\end{align}

We are interested in the limit $k_1/k_F\rightarrow \infty$. To avoid numerical integration up to infinity we use that for the region $p \gg k_2$ we can approximate $p\approx q$ and get
\begin{align}
I_{>}=-\frac{D}{4\pi} \frac{\epsilon_F}{\epsilon_1}  \int_{\Lambda}^{\infty} d\bar{p}\int_{0}^{1} d\bar{k_2}\int_{\bar{p}}^{2\bar{k_2}+\bar{p}} d\bar{q}  \sqrt{2\bar{p}} \sqrt{2\bar{k_2}+\bar{p}-\bar{q}}\frac{1}{\bar{p}^2} \frac{1}{\bar{k_2}}=-\frac{D}{4\pi} \frac{8}{3}\frac{2}{3}\frac{\epsilon_F}{\epsilon_1}  \int_{\Lambda}^{\infty} d\bar{p} \frac{1}{\bar{p}^{3/2}},
\label{eq:ederivlinlargeromega2}
\end{align}
where $\Lambda$ is a cutoff that ensures that the approximation $p \gg k_2$ is valid.
The remaining part of the integral we cannot approximate further and we have to integrate
\begin{align}
I_{<}=-\frac{D}{4\pi} \frac{\epsilon_F}{\epsilon_1}  \int_{1}^{\Lambda} d\bar{p}\int_{0}^{1} d\bar{k_2}\int_{\bar{p}}^{2\bar{k_2}+\bar{p}} d\bar{q}  \sqrt{2\bar{p}} \sqrt{2\bar{k_2}+\bar{p}-\bar{q}}\left(\frac{\bar{p}}{\bar{q}}\right)\frac{1}{(\bar{q}+\alpha)^2} \frac{(2\bar{k_2}+\bar{p})}{\bar{k_2}(\bar{k_2}+\bar{p})}
\label{eq:ederivlinlargeromega3}
\end{align}
numerically.
For $\Lambda=10$ we get $I_{>}=-\frac{16}{9}\sqrt{2/5} D/(4\pi) (\epsilon_F/\epsilon_1)$ and $I_{<}\approx -1.93 D/(4\pi) (\epsilon_F/\epsilon_1)$ and for $\Lambda=100$ we get $I_{>}=-\frac{16}{9}\frac{1}{5} D/(4\pi) (\epsilon_F/\epsilon_1)$ and $I_{<}\approx - 2.68 D/(4\pi) (\epsilon_F/\epsilon_1)$. Both cutoffs give us the same final result of
\begin{align}
 \frac{d \bm{j}^{e}_d}{dt}+\frac{d j^{e}_{ex}}{dt}\approx 4.2 \frac{D}{4\pi} \frac{\epsilon_F}{\epsilon_1} \hat{j_0} \approx 0.3 e v_F \alpha^2 \frac{\epsilon_F}{\hbar} \hat{j_0}.
\label{eq:ederivlin6}
\end{align}

An analogous calculation for the interference term shows that $\frac{d \bm{j}^{e}_{\textnormal{inter}}}{dt}\sim -D /(4\pi)(\epsilon_F/\epsilon_1)^{3/2}$, which is of higher order.

It can be easily shown that to lowest order in $\epsilon_F/\epsilon_1$, $(d j^{e}_d/dt) + (d j^{e}_{ex}/dt)$ and $(d j^{h}_{d}/dt)$ just differ by a sign. In Fig.~\ref{fig:holeprocess}(b), labeling the initial states by $\bm{k}_1$, $\bm{k}_2'$ and the final states by $\bm{k}_1'$, $\bm{k}_2$, i.e., switching $\bm{k}_2 \leftrightarrow \bm{k}_2'$, and making use of the approximations \eqref{eq:approx1} and \eqref{eq:approx2}, the direct term of the hole current can be written as

\begin{align}
\frac{d \bm{j}_{d}^{h}}{dt}&\approx 2 e \frac{1}{4 L^4}\frac{2 \pi}{\hbar} \left(\frac{e^2}{2 \epsilon_0\epsilon}\right)^2\sum_{\bm{k}_2,\bm{k}_2',\bm{k}_1'}\delta(\epsilon_{\bm{k}_1}+\epsilon_{\bm{k}_2'}-\epsilon_{\bm{k}_1'}-\epsilon_{\bm{k}_2}) \left(\hat{k}_2-\hat{k}_2'\right) \left\vert \frac{\langle \bm{k}_2',+\vert\bm{k}_2,+\rangle}{\vert \bm{k}_1-\bm{k}_1'\vert+q_{TF}}\right\vert^2 \theta(\epsilon_{\bm{k}_2'}-\epsilon_F)\theta(\epsilon_F-\epsilon_{\bm{k}_2}),
\label{eq:hderivlin1}
\end{align}
where $\bm{k}_2'=\bm{k}_2-\bm{q}$. The transformation $\phi_2\rightarrow\phi_2+\pi$, leads to $\bm{k}_2\rightarrow -\bm{k}_2$ and $\bm{k}_2-\bm{q}\rightarrow -(\bm{k}_2+\bm{q})$. With $\vert- \bm{k},\pm\rangle=\vert\bm{k},\mp\rangle$ and $\vert \langle \bm{k},+ \vert\bm{k}',+\rangle \vert^2=\vert \langle \bm{k},- \vert\bm{k}',-\rangle \vert^2$, we find to lowest order that 
\begin{equation}
 \frac{d \bm{j}_{d}^{h}}{dt}=-\left(\frac{d \bm{j}_{d}^{e}}{dt}+\frac{d \bm{j}_{ex}^{e}}{dt}\right).
\end{equation}

Analogous calculations to the one above for $d j^{e}_d/dt$ lead to
\begin{eqnarray}
 \frac{d \bm{j}_{\textnormal{inter}}^{e}}{dt} &\sim& \frac{D}{4\pi} \left(\frac{\epsilon_F}{\epsilon_1}\right)^{3/2} \hat{j_0} \nonumber\\
 \frac{d \bm{j}_{\textnormal{inter}}^{h}}{dt} &\sim& -\frac{D}{4\pi} \left(\frac{\epsilon_F}{\epsilon_1}\right)^{3/2} \hat{j_0} \nonumber\\
 \frac{d \bm{j}_{\textnormal{inter}}^{e}}{dt} +\frac{d \bm{j}_{\textnormal{inter}}^{h}}{dt} &\sim& -\frac{D}{4\pi} \left(\frac{\epsilon_F}{\epsilon_1}\right)^{2} \hat{j_0} \nonumber\\
 \frac{d \bm{j}_{d}^{e}}{dt}+\frac{d \bm{j}_{ex}^{e}}{dt} +\frac{d \bm{j}_{d}^{h}}{dt} &\sim& -\frac{D}{4\pi} \left(\frac{\epsilon_F}{\epsilon_1}\right)^{5/2} \hat{j_0} \nonumber \\
\frac{d \bm{j}_{ex}^{h}}{dt} &\approx& -\frac{D}{9} \left(\frac{\epsilon_F}{\epsilon_1}\right)^{3/2} \hat{j_0}.
\end{eqnarray}
Calculating the rate of change of the total current, we then get for the asymptotic behavior in the limit $\epsilon_F \ll \epsilon_1$,
\begin{align}
 \frac{d \bm{j}^{\textnormal{(tot)}}}{dt}&= \frac{d \bm{j}_{d}^{e}}{dt}+\frac{d \bm{j}_{ex}^{e}}{dt} +\frac{d \bm{j}_{d}^{h}}{dt} -\left(\frac{d \bm{j}_{\textnormal{inter}}^{e}}{dt} +\frac{d \bm{j}_{\textnormal{inter}}^{h}}{dt} \right)+ \frac{d \bm{j}_{ex}^{h}}{dt} \nonumber\\
&\approx  \frac{d \bm{j}_{ex}^{h}}{dt} +\mathcal{O}\left[\left(\frac{\epsilon_F}{\epsilon_1}\right)^{2} \right] \nonumber\\
&\approx -\frac{D}{9} \left(\frac{\epsilon_F}{\epsilon_1}\right)^{3/2} \hat{j_0} +\mathcal{O}\left[\left(\frac{\epsilon_F}{\epsilon_1}\right)^{2} \right].
\end{align}

\section{E) Definition of the mean change of the current per electron scattering event}

The mean change in current per scattering event is defined by
\begin{equation}
 \frac{\langle \Delta j\rangle}{j_0}=\frac{1}{j_0\Gamma}\frac{dj}{dt}.
\end{equation}
While $dj/dt$ is always well defined, $\Gamma$ diverges for a perfectly linear dispersion because the phase space for collinear scattering becomes infinite. We regularize this by introducing a physical and commonly used particle-hole asymmetry, such that $\epsilon_{\bm{k}}=\xi k^2 \pm v_F k$. When calculating the now well defined $\Gamma$ we have to take into account all allowed scattering processes. Processes that are collinear and can be neglected in the calculation of the rate of change of the current have to be included in $\Gamma$.




\begin{thebibliography}{36}
\expandafter\ifx\csname natexlab\endcsname\relax\def\natexlab#1{#1}\fi
\expandafter\ifx\csname bibnamefont\endcsname\relax
  \def\bibnamefont#1{#1}\fi
\expandafter\ifx\csname bibfnamefont\endcsname\relax
  \def\bibfnamefont#1{#1}\fi
\expandafter\ifx\csname citenamefont\endcsname\relax
  \def\citenamefont#1{#1}\fi
\expandafter\ifx\csname url\endcsname\relax
  \def\url#1{\texttt{#1}}\fi
\expandafter\ifx\csname urlprefix\endcsname\relax\def\urlprefix{URL }\fi
\providecommand{\bibinfo}[2]{#2}
\providecommand{\eprint}[2][]{\url{#2}}

\bibitem[{\citenamefont{Novoselov et~al.}(2005)\citenamefont{Novoselov, Jiang,
  Schedin, Booth, Khotkevich, Morozov, and Geim}}]{novoselov2005}
\bibinfo{author}{\bibfnamefont{K.~S.} \bibnamefont{Novoselov}},
  \bibinfo{author}{\bibfnamefont{D.}~\bibnamefont{Jiang}},
  \bibinfo{author}{\bibfnamefont{F.}~\bibnamefont{Schedin}},
  \bibinfo{author}{\bibfnamefont{T.~J.} \bibnamefont{Booth}},
  \bibinfo{author}{\bibfnamefont{V.~V.} \bibnamefont{Khotkevich}},
  \bibinfo{author}{\bibfnamefont{S.~V.} \bibnamefont{Morozov}},
  \bibnamefont{and} \bibinfo{author}{\bibfnamefont{A.~K.} \bibnamefont{Geim}},
  \bibinfo{journal}{Proceedings of the National Academy of Sciences of the
  United States of America} \textbf{\bibinfo{volume}{102}},
  \bibinfo{pages}{10451} (\bibinfo{year}{2005}).

\bibitem[{\citenamefont{Kane and Mele}(2005)}]{kane2005}
\bibinfo{author}{\bibfnamefont{C.~L.} \bibnamefont{Kane}} \bibnamefont{and}
  \bibinfo{author}{\bibfnamefont{E.~J.} \bibnamefont{Mele}},
  \bibinfo{journal}{Phys. Rev. Lett.} \textbf{\bibinfo{volume}{95}},
  \bibinfo{pages}{146802} (\bibinfo{year}{2005}).

\bibitem[{\citenamefont{K\"onig et~al.}(2007)\citenamefont{K\"onig, Wiedmann,
  Br\"une, Roth, Buhmann, Molenkamp, Qi, and Zhang}}]{konig2007}
\bibinfo{author}{\bibfnamefont{M.}~\bibnamefont{K\"onig}},
  \bibinfo{author}{\bibfnamefont{S.}~\bibnamefont{Wiedmann}},
  \bibinfo{author}{\bibfnamefont{C.}~\bibnamefont{Br\"une}},
  \bibinfo{author}{\bibfnamefont{A.}~\bibnamefont{Roth}},
  \bibinfo{author}{\bibfnamefont{H.}~\bibnamefont{Buhmann}},
  \bibinfo{author}{\bibfnamefont{L.~W.} \bibnamefont{Molenkamp}},
  \bibinfo{author}{\bibfnamefont{X.-L.} \bibnamefont{Qi}}, \bibnamefont{and}
  \bibinfo{author}{\bibfnamefont{S.-C.} \bibnamefont{Zhang}},
  \bibinfo{journal}{Science} \textbf{\bibinfo{volume}{318}},
  \bibinfo{pages}{766} (\bibinfo{year}{2007}).

\bibitem[{\citenamefont{Castro~Neto et~al.}(2009)\citenamefont{Castro~Neto,
  Guinea, Peres, Novoselov, and Geim}}]{castro2009}
\bibinfo{author}{\bibfnamefont{A.~H.} \bibnamefont{Castro~Neto}},
  \bibinfo{author}{\bibfnamefont{F.}~\bibnamefont{Guinea}},
  \bibinfo{author}{\bibfnamefont{N.~M.~R.} \bibnamefont{Peres}},
  \bibinfo{author}{\bibfnamefont{K.~S.} \bibnamefont{Novoselov}},
  \bibnamefont{and} \bibinfo{author}{\bibfnamefont{A.~K.} \bibnamefont{Geim}},
  \bibinfo{journal}{Rev. Mod. Phys.} \textbf{\bibinfo{volume}{81}},
  \bibinfo{pages}{109} (\bibinfo{year}{2009}).

\bibitem[{\citenamefont{Fu and Kane}(2007)}]{fu2007}
\bibinfo{author}{\bibfnamefont{L.}~\bibnamefont{Fu}} \bibnamefont{and}
  \bibinfo{author}{\bibfnamefont{C.~L.} \bibnamefont{Kane}},
  \bibinfo{journal}{Phys. Rev. B} \textbf{\bibinfo{volume}{76}},
  \bibinfo{pages}{045302} (\bibinfo{year}{2007}).

\bibitem[{\citenamefont{Hsieh et~al.}(2008)\citenamefont{Hsieh, Qian, Wray,
  Xia, Hor, Cava, and Hasan}}]{hsieh2008}
\bibinfo{author}{\bibfnamefont{D.}~\bibnamefont{Hsieh}},
  \bibinfo{author}{\bibfnamefont{D.}~\bibnamefont{Qian}},
  \bibinfo{author}{\bibfnamefont{L.}~\bibnamefont{Wray}},
  \bibinfo{author}{\bibfnamefont{Y.}~\bibnamefont{Xia}},
  \bibinfo{author}{\bibfnamefont{Y.~S.} \bibnamefont{Hor}},
  \bibinfo{author}{\bibfnamefont{R.~J.} \bibnamefont{Cava}}, \bibnamefont{and}
  \bibinfo{author}{\bibfnamefont{M.~Z.} \bibnamefont{Hasan}},
  \bibinfo{journal}{Nature} \textbf{\bibinfo{volume}{452}},
  \bibinfo{pages}{970} (\bibinfo{year}{2008}).

\bibitem[{\citenamefont{Hasan and Kane}(2010)}]{hasan2010}
\bibinfo{author}{\bibfnamefont{M.~Z.} \bibnamefont{Hasan}} \bibnamefont{and}
  \bibinfo{author}{\bibfnamefont{C.~L.} \bibnamefont{Kane}},
  \bibinfo{journal}{Rev. Mod. Phys.} \textbf{\bibinfo{volume}{82}},
  \bibinfo{pages}{3045} (\bibinfo{year}{2010}).

\bibitem[{\citenamefont{Qi and Zhang}(2011)}]{qi2011}
\bibinfo{author}{\bibfnamefont{X.-L.} \bibnamefont{Qi}} \bibnamefont{and}
  \bibinfo{author}{\bibfnamefont{S.-C.} \bibnamefont{Zhang}},
  \bibinfo{journal}{Rev. Mod. Phys.} \textbf{\bibinfo{volume}{83}},
  \bibinfo{pages}{1057} (\bibinfo{year}{2011}).

\bibitem[{\citenamefont{MacDonald}(2012)}]{macdonald2012}
\bibinfo{author}{\bibfnamefont{A.~H.} \bibnamefont{MacDonald}},
  \bibinfo{journal}{Nat. Mater.} \textbf{\bibinfo{volume}{11}},
  \bibinfo{pages}{409} (\bibinfo{year}{2012}).

\bibitem[{\citenamefont{Bonaccorso et~al.}(2010)\citenamefont{Bonaccorso, Sun,
  Hasan, and Ferrari}}]{bonaccorso2010}
\bibinfo{author}{\bibfnamefont{F.}~\bibnamefont{Bonaccorso}},
  \bibinfo{author}{\bibfnamefont{Z.}~\bibnamefont{Sun}},
  \bibinfo{author}{\bibfnamefont{T.}~\bibnamefont{Hasan}}, \bibnamefont{and}
  \bibinfo{author}{\bibfnamefont{A.~C.} \bibnamefont{Ferrari}},
  \bibinfo{journal}{Nature Photonics} \textbf{\bibinfo{volume}{4}},
  \bibinfo{pages}{611} (\bibinfo{year}{2010}).

\bibitem[{\citenamefont{Kong and Cui}(2010)}]{kong2010}
\bibinfo{author}{\bibfnamefont{D.}~\bibnamefont{Kong}} \bibnamefont{and}
  \bibinfo{author}{\bibfnamefont{Y.}~\bibnamefont{Cui}},
  \bibinfo{journal}{Nature Chemistry} \textbf{\bibinfo{volume}{3}},
  \bibinfo{pages}{845} (\bibinfo{year}{2010}).

\bibitem[{\citenamefont{Park et~al.}(2009)\citenamefont{Park, Ahn, and
  Ruiz-Vargas}}]{park2009}
\bibinfo{author}{\bibfnamefont{J.}~\bibnamefont{Park}},
  \bibinfo{author}{\bibfnamefont{Y.~H.} \bibnamefont{Ahn}}, \bibnamefont{and}
  \bibinfo{author}{\bibfnamefont{C.}~\bibnamefont{Ruiz-Vargas}},
  \bibinfo{journal}{Nano Letters} \textbf{\bibinfo{volume}{9}},
  \bibinfo{pages}{1742} (\bibinfo{year}{2009}).

\bibitem[{\citenamefont{{J.W. McIver} et~al.}(2011)\citenamefont{{J.W. McIver},
  {D. Hsieh}, {H. Steinberg}, {P. Jarillo-Herrero}, and {N.
  Gedik}}}]{mciver2011}
\bibinfo{author}{\bibnamefont{{J.W. McIver}}},
  \bibinfo{author}{\bibnamefont{{D. Hsieh}}}, \bibinfo{author}{\bibnamefont{{H.
  Steinberg}}}, \bibinfo{author}{\bibnamefont{{P. Jarillo-Herrero}}},
  \bibnamefont{and} \bibinfo{author}{\bibnamefont{{N. Gedik}}},
  \bibinfo{journal}{Nat. Nanotech.} \textbf{\bibinfo{volume}{7}},
  \bibinfo{pages}{96} (\bibinfo{year}{2011}).

\bibitem[{\citenamefont{Kastl et~al.}(2012)\citenamefont{Kastl, Guan, He, Wu,
  and Li}}]{kastl2012}
\bibinfo{author}{\bibfnamefont{C.}~\bibnamefont{Kastl}},
  \bibinfo{author}{\bibfnamefont{T.}~\bibnamefont{Guan}},
  \bibinfo{author}{\bibfnamefont{X.~Y.} \bibnamefont{He}},
  \bibinfo{author}{\bibfnamefont{K.~H.} \bibnamefont{Wu}}, \bibnamefont{and}
  \bibinfo{author}{\bibfnamefont{Y.~Q.} \bibnamefont{Li}},
  \bibinfo{journal}{Appl. Phys. Lett.} \textbf{\bibinfo{volume}{101}},
  \bibinfo{pages}{251110} (\bibinfo{year}{2012}).

\bibitem[{\citenamefont{Sun et~al.}(2012)\citenamefont{Sun, Divin, Mihnev,
  Winzer, Malic, Knorr, Sipe, Berger, de~Heer, First et~al.}}]{sun2012}
\bibinfo{author}{\bibfnamefont{D.}~\bibnamefont{Sun}},
  \bibinfo{author}{\bibfnamefont{C.}~\bibnamefont{Divin}},
  \bibinfo{author}{\bibfnamefont{M.}~\bibnamefont{Mihnev}},
  \bibinfo{author}{\bibfnamefont{T.}~\bibnamefont{Winzer}},
  \bibinfo{author}{\bibfnamefont{E.}~\bibnamefont{Malic}},
  \bibinfo{author}{\bibfnamefont{A.}~\bibnamefont{Knorr}},
  \bibinfo{author}{\bibfnamefont{J.~E.} \bibnamefont{Sipe}},
  \bibinfo{author}{\bibfnamefont{C.}~\bibnamefont{Berger}},
  \bibinfo{author}{\bibfnamefont{W.~A.} \bibnamefont{de~Heer}},
  \bibinfo{author}{\bibfnamefont{P.~N.} \bibnamefont{First}},
  \bibnamefont{et~al.}, \bibinfo{journal}{New Journal of Physics}
  \textbf{\bibinfo{volume}{14}}, \bibinfo{pages}{105012}
  (\bibinfo{year}{2012}).

\bibitem[{\citenamefont{Freitag et~al.}(2013)\citenamefont{Freitag, Low, Zhu,
  Yan, Xia, and Avouris}}]{freitag2013}
\bibinfo{author}{\bibfnamefont{M.}~\bibnamefont{Freitag}},
  \bibinfo{author}{\bibfnamefont{T.}~\bibnamefont{Low}},
  \bibinfo{author}{\bibfnamefont{W.}~\bibnamefont{Zhu}},
  \bibinfo{author}{\bibfnamefont{H.}~\bibnamefont{Yan}},
  \bibinfo{author}{\bibfnamefont{F.}~\bibnamefont{Xia}}, \bibnamefont{and}
  \bibinfo{author}{\bibfnamefont{P.}~\bibnamefont{Avouris}},
  \bibinfo{journal}{Nat. Comm.} \textbf{\bibinfo{volume}{4}}
  (\bibinfo{year}{2013}).

\bibitem[{\citenamefont{Junck et~al.}(2013)\citenamefont{Junck, Refael, and von
  Oppen}}]{junck2013}
\bibinfo{author}{\bibfnamefont{A.}~\bibnamefont{Junck}},
  \bibinfo{author}{\bibfnamefont{G.}~\bibnamefont{Refael}}, \bibnamefont{and}
  \bibinfo{author}{\bibfnamefont{F.}~\bibnamefont{von Oppen}},
  \bibinfo{journal}{Phys. Rev. B} \textbf{\bibinfo{volume}{88}},
  \bibinfo{pages}{075144} (\bibinfo{year}{2013}).

\bibitem[{\citenamefont{Cheianov and Fal'ko}(2006)}]{cheianov2006}
\bibinfo{author}{\bibfnamefont{V.~V.} \bibnamefont{Cheianov}} \bibnamefont{and}
  \bibinfo{author}{\bibfnamefont{V.~I.} \bibnamefont{Fal'ko}},
  \bibinfo{journal}{Phys. Rev. Lett.} \textbf{\bibinfo{volume}{97}},
  \bibinfo{pages}{226801} (\bibinfo{year}{2006}).

\bibitem[{\citenamefont{Butscher et~al.}(2007)\citenamefont{Butscher, Milde,
  Hirtschulz, Malic, and Knorr}}]{butscher2007}
\bibinfo{author}{\bibfnamefont{S.}~\bibnamefont{Butscher}},
  \bibinfo{author}{\bibfnamefont{F.}~\bibnamefont{Milde}},
  \bibinfo{author}{\bibfnamefont{M.}~\bibnamefont{Hirtschulz}},
  \bibinfo{author}{\bibfnamefont{E.}~\bibnamefont{Malic}}, \bibnamefont{and}
  \bibinfo{author}{\bibfnamefont{A.}~\bibnamefont{Knorr}},
  \bibinfo{journal}{Applied Physics Letters} \textbf{\bibinfo{volume}{91}},
  \bibinfo{eid}{203103} (\bibinfo{year}{2007}).

\bibitem[{\citenamefont{Stauber et~al.}(2007)\citenamefont{Stauber, Peres, and
  Guinea}}]{stauber2007}
\bibinfo{author}{\bibfnamefont{T.}~\bibnamefont{Stauber}},
  \bibinfo{author}{\bibfnamefont{N.~M.~R.} \bibnamefont{Peres}},
  \bibnamefont{and} \bibinfo{author}{\bibfnamefont{F.}~\bibnamefont{Guinea}},
  \bibinfo{journal}{Phys. Rev. B} \textbf{\bibinfo{volume}{76}},
  \bibinfo{pages}{205423} (\bibinfo{year}{2007}).

\bibitem[{\citenamefont{Tse and Das~Sarma}(2009)}]{tse2009}
\bibinfo{author}{\bibfnamefont{W.-K.} \bibnamefont{Tse}} \bibnamefont{and}
  \bibinfo{author}{\bibfnamefont{S.}~\bibnamefont{Das~Sarma}},
  \bibinfo{journal}{Phys. Rev. B} \textbf{\bibinfo{volume}{79}},
  \bibinfo{pages}{235406} (\bibinfo{year}{2009}).

\bibitem[{\citenamefont{Winzer et~al.}(2010)\citenamefont{Winzer, Knorr, and
  Malic}}]{winzer2010}
\bibinfo{author}{\bibfnamefont{T.}~\bibnamefont{Winzer}},
  \bibinfo{author}{\bibfnamefont{A.}~\bibnamefont{Knorr}}, \bibnamefont{and}
  \bibinfo{author}{\bibfnamefont{E.}~\bibnamefont{Malic}},
  \bibinfo{journal}{Nano Letters} \textbf{\bibinfo{volume}{10}},
  \bibinfo{pages}{4839} (\bibinfo{year}{2010}).

\bibitem[{\citenamefont{Kim et~al.}(2011)\citenamefont{Kim, Perebeinos, and
  Avouris}}]{kim2011}
\bibinfo{author}{\bibfnamefont{R.}~\bibnamefont{Kim}},
  \bibinfo{author}{\bibfnamefont{V.}~\bibnamefont{Perebeinos}},
  \bibnamefont{and} \bibinfo{author}{\bibfnamefont{P.}~\bibnamefont{Avouris}},
  \bibinfo{journal}{Phys. Rev. B} \textbf{\bibinfo{volume}{84}},
  \bibinfo{pages}{075449} (\bibinfo{year}{2011}).

\bibitem[{\citenamefont{Song et~al.}(2013)\citenamefont{Song, Tielrooij,
  Koppens, and Levitov}}]{song2013}
\bibinfo{author}{\bibfnamefont{J.~C.~W.} \bibnamefont{Song}},
  \bibinfo{author}{\bibfnamefont{K.~J.} \bibnamefont{Tielrooij}},
  \bibinfo{author}{\bibfnamefont{F.~H.~L.} \bibnamefont{Koppens}},
  \bibnamefont{and} \bibinfo{author}{\bibfnamefont{L.~S.}
  \bibnamefont{Levitov}}, \bibinfo{journal}{Phys. Rev. B}
  \textbf{\bibinfo{volume}{87}}, \bibinfo{pages}{155429}
  (\bibinfo{year}{2013}).

\bibitem[{\citenamefont{Hsieh et~al.}(2011)\citenamefont{Hsieh, Mahmood,
  McIver, Gardner, Lee, and Gedik}}]{hsieh2007}
\bibinfo{author}{\bibfnamefont{D.}~\bibnamefont{Hsieh}},
  \bibinfo{author}{\bibfnamefont{F.}~\bibnamefont{Mahmood}},
  \bibinfo{author}{\bibfnamefont{J.~W.} \bibnamefont{McIver}},
  \bibinfo{author}{\bibfnamefont{D.~R.} \bibnamefont{Gardner}},
  \bibinfo{author}{\bibfnamefont{Y.~S.} \bibnamefont{Lee}}, \bibnamefont{and}
  \bibinfo{author}{\bibfnamefont{N.}~\bibnamefont{Gedik}},
  \bibinfo{journal}{Phys. Rev. Lett.} \textbf{\bibinfo{volume}{107}},
  \bibinfo{pages}{077401} (\bibinfo{year}{2011}).

\bibitem[{\citenamefont{Kumar et~al.}(2011)\citenamefont{Kumar, Ruzicka, Butch,
  Syers, Kirshenbaum, Paglione, and Zhao}}]{kumar2011}
\bibinfo{author}{\bibfnamefont{N.}~\bibnamefont{Kumar}},
  \bibinfo{author}{\bibfnamefont{B.~A.} \bibnamefont{Ruzicka}},
  \bibinfo{author}{\bibfnamefont{N.~P.} \bibnamefont{Butch}},
  \bibinfo{author}{\bibfnamefont{P.}~\bibnamefont{Syers}},
  \bibinfo{author}{\bibfnamefont{K.}~\bibnamefont{Kirshenbaum}},
  \bibinfo{author}{\bibfnamefont{J.}~\bibnamefont{Paglione}}, \bibnamefont{and}
  \bibinfo{author}{\bibfnamefont{H.}~\bibnamefont{Zhao}},
  \bibinfo{journal}{Phys. Rev. B} \textbf{\bibinfo{volume}{83}},
  \bibinfo{pages}{235306} (\bibinfo{year}{2011}).

\bibitem[{\citenamefont{Sobota et~al.}(2012)\citenamefont{Sobota, Yang,
  Analytis, Chen, Fisher, Kirchmann, and Shen}}]{sobota2012}
\bibinfo{author}{\bibfnamefont{J.~A.} \bibnamefont{Sobota}},
  \bibinfo{author}{\bibfnamefont{S.}~\bibnamefont{Yang}},
  \bibinfo{author}{\bibfnamefont{J.~G.} \bibnamefont{Analytis}},
  \bibinfo{author}{\bibfnamefont{Y.~L.} \bibnamefont{Chen}},
  \bibinfo{author}{\bibfnamefont{I.~R.} \bibnamefont{Fisher}},
  \bibinfo{author}{\bibfnamefont{P.~S.} \bibnamefont{Kirchmann}},
  \bibnamefont{and} \bibinfo{author}{\bibfnamefont{Z.-X.} \bibnamefont{Shen}},
  \bibinfo{journal}{Phys. Rev. Lett.} \textbf{\bibinfo{volume}{108}},
  \bibinfo{pages}{117403} (\bibinfo{year}{2012}).

\bibitem[{\citenamefont{Hajlaoui et~al.}(2012)\citenamefont{Hajlaoui,
  Papalazarou, Mauchain, Lantz, Moisan, Boschetto, Jiang, Miotkowski, Chen,
  Taleb-Ibrahimi et~al.}}]{hajlaoui2012}
\bibinfo{author}{\bibfnamefont{M.}~\bibnamefont{Hajlaoui}},
  \bibinfo{author}{\bibfnamefont{E.}~\bibnamefont{Papalazarou}},
  \bibinfo{author}{\bibfnamefont{J.}~\bibnamefont{Mauchain}},
  \bibinfo{author}{\bibfnamefont{G.}~\bibnamefont{Lantz}},
  \bibinfo{author}{\bibfnamefont{N.}~\bibnamefont{Moisan}},
  \bibinfo{author}{\bibfnamefont{D.}~\bibnamefont{Boschetto}},
  \bibinfo{author}{\bibfnamefont{Z.}~\bibnamefont{Jiang}},
  \bibinfo{author}{\bibfnamefont{I.}~\bibnamefont{Miotkowski}},
  \bibinfo{author}{\bibfnamefont{Y.~P.} \bibnamefont{Chen}},
  \bibinfo{author}{\bibfnamefont{A.}~\bibnamefont{Taleb-Ibrahimi}},
  \bibnamefont{et~al.}, \bibinfo{journal}{Nano Letters}
  \textbf{\bibinfo{volume}{12}}, \bibinfo{pages}{3532} (\bibinfo{year}{2012}).

\bibitem[{\citenamefont{Gierz et~al.}(2013)\citenamefont{Gierz, Petersen,
  Mitrano, Cacho, Turcu, Springate, St\"or, Axel, Starke, and
  Cavalleri}}]{gierz2013}
\bibinfo{author}{\bibfnamefont{I.}~\bibnamefont{Gierz}},
  \bibinfo{author}{\bibfnamefont{J.~C.} \bibnamefont{Petersen}},
  \bibinfo{author}{\bibfnamefont{M.}~\bibnamefont{Mitrano}},
  \bibinfo{author}{\bibfnamefont{C.}~\bibnamefont{Cacho}},
  \bibinfo{author}{\bibfnamefont{I.~C.~E.} \bibnamefont{Turcu}},
  \bibinfo{author}{\bibfnamefont{E.}~\bibnamefont{Springate}},
  \bibinfo{author}{\bibfnamefont{A.}~\bibnamefont{St\"or}},
  \bibinfo{author}{\bibfnamefont{K.}~\bibnamefont{Axel}},
  \bibinfo{author}{\bibfnamefont{U.}~\bibnamefont{Starke}}, \bibnamefont{and}
  \bibinfo{author}{\bibfnamefont{A.}~\bibnamefont{Cavalleri}},
  \bibinfo{journal}{Nature Mater.} \textbf{\bibinfo{volume}{12}},
  \bibinfo{pages}{1119} (\bibinfo{year}{2013}).

\bibitem[{\citenamefont{Liu et~al.}(2010)\citenamefont{Liu, Qi, Zhang, Dai,
  Fang, and Zhang}}]{liu2010}
\bibinfo{author}{\bibfnamefont{C.}~\bibnamefont{Liu}},
  \bibinfo{author}{\bibfnamefont{X.}~\bibnamefont{Qi}},
  \bibinfo{author}{\bibfnamefont{H.}~\bibnamefont{Zhang}},
  \bibinfo{author}{\bibfnamefont{X.}~\bibnamefont{Dai}},
  \bibinfo{author}{\bibfnamefont{Z.}~\bibnamefont{Fang}}, \bibnamefont{and}
  \bibinfo{author}{\bibfnamefont{S.}~\bibnamefont{Zhang}},
  \bibinfo{journal}{Phys. Rev. B} \textbf{\bibinfo{volume}{82}},
  \bibinfo{pages}{045122} (\bibinfo{year}{2010}).

\bibitem[{\citenamefont{Greenaway and Harbeke}(1965)}]{greenaway1965}
\bibinfo{author}{\bibfnamefont{D.}~\bibnamefont{Greenaway}} \bibnamefont{and}
  \bibinfo{author}{\bibfnamefont{G.}~\bibnamefont{Harbeke}},
  \bibinfo{journal}{J. Phys. Chem. Solids} \textbf{\bibinfo{volume}{26}},
  \bibinfo{pages}{1585 } (\bibinfo{year}{1965}).

\bibitem[{\citenamefont{Sandomirsky et~al.}(2001)\citenamefont{Sandomirsky,
  Butenko, Levin, and Schlesinger}}]{sandomirsky2001}
\bibinfo{author}{\bibfnamefont{V.}~\bibnamefont{Sandomirsky}},
  \bibinfo{author}{\bibfnamefont{A.~V.} \bibnamefont{Butenko}},
  \bibinfo{author}{\bibfnamefont{R.}~\bibnamefont{Levin}}, \bibnamefont{and}
  \bibinfo{author}{\bibfnamefont{Y.}~\bibnamefont{Schlesinger}},
  \bibinfo{journal}{J. Appl. Phys.} \textbf{\bibinfo{volume}{90}},
  \bibinfo{pages}{2370} (\bibinfo{year}{2001}).

\bibitem[]{suppmat}
\bibinfo{suppmat}{See Supplemental Material for details}.

\bibitem[]{note}
\bibinfo{note}{The total scattering rate $\Gamma $ actually diverges for
  a perfectly linear dispersion. As this divergence is due to almost collinear
  scattering, it only affects the total scattering rate. The rate of change of
  the current remains always well defined. In Fig.~\ref{fig:meancurrkF}, the
  divergence of $\Gamma $ is regularized by the physical particle-hole
  asymmetry. For perfectly linear dispersion, the divergence could also be
  regularized by dynamic screening within the random phase approximation \cite{song2013}.}

\bibitem[{\citenamefont{Tani et~al.}(2012)\citenamefont{Tani, Blanchard, and
  Tanaka}}]{tani2012}
\bibinfo{author}{\bibfnamefont{S.}~\bibnamefont{Tani}},
  \bibinfo{author}{\bibfnamefont{F.}~\bibnamefont{Blanchard}},
  \bibnamefont{and} \bibinfo{author}{\bibfnamefont{K.}~\bibnamefont{Tanaka}},
  \bibinfo{journal}{Phys. Rev. Lett.} \textbf{\bibinfo{volume}{109}},
  \bibinfo{pages}{166603} (\bibinfo{year}{2012}).

\bibitem[]{SongPrivate}
\bibinfo{note}{J.~C.~W. Song (private communication)}.

\end{thebibliography}
\end{document}